\documentclass[sigconf,balance=false]{acmart}

\usepackage{popets}
\setcopyright{popets}
\copyrightyear{2024}
\acmYear{2024}
\acmVolume{2024}
\acmNumber{1}
\acmDOI{XXXXXXX.XXXXXXX}
\acmISBN{}
\acmConference{Proceedings on Privacy Enhancing Technologies}
\acmSubmissionID{\#5}
\usepackage{enumitem}
\usepackage{tcolorbox}
\usepackage[all]{nowidow}
\usepackage{siunitx}
\usepackage[linesnumbered,ruled,vlined]{algorithm2e}
\usepackage{multirow}
\usepackage{amsmath}
\usepackage{caption}
\usepackage{subcaption}
\usepackage{colortbl}
\usepackage[subtle]{savetrees}

\usepackage{diagbox, eqparbox, hhline}
\DeclareMathOperator{\E}{\mathbb{E}}

\newcommand{\sandbox}{\texttt{The Privacy Sandbox}}
\newcommand{\floc}{\texttt{FLoC}}
\newcommand{\topics}{\texttt{Topics}}
\newcommand{\browsingTopics}{\textit{<browsingTopics()>}}
\newcommand{\unknownTopic}{\textit{Unknown}}

\newcommand{\eg}{e.g.,}
\newcommand{\ie}{i.e.,}
\newcommand{\etc}{etc.}

\newcommand{\goal}[1]{(\textbf{G{#1}})}

\newcommand{\shortsection}[2][.]{%  % better sub-sub-sub-sections (with a lighter bold)
\vspace{1mm}\noindent\textbf{#2#1}}

\definecolor{Gray}{gray}{0.9}

\begin{document}

\title[]{Interest-disclosing Mechanisms for Advertising are\\ Privacy-\textit{Exposing} (not Preserving)}

%%%%%%%%%%%%%%%% Authors' Info %%%%%%%%%%%%%%%%%

\author{Yohan Beugin}
\orcid{0000-0003-0991-7926}
\affiliation{%
  \institution{University of Wisconsin-Madison}
  \city{Madison}
  \state{Wisconsin}
  \country{USA}}
\email{ybeugin@cs.wisc.edu}

\author{Patrick McDaniel}
\orcid{0000-0003-2091-7484}
\affiliation{%
  \institution{University of Wisconsin-Madison}
  \city{Madison}
  \state{Wisconsin}
  \country{USA}}
\email{mcdaniel@cs.wisc.edu}


% \renewcommand{\shortauthors}{Beugin et al.}

\begin{abstract}
Today, targeted online advertising relies on unique identifiers assigned to
users through third-party cookies--a practice at odds with user privacy. While
the web and advertising communities have proposed solutions that we refer to as
interest-disclosing mechanisms, including Google's \topics{} API, an independent
analysis of these proposals in realistic scenarios has yet to be performed. In
this paper, we attempt to validate the privacy (\ie{} preventing unique
identification) and utility (\ie{} enabling ad targeting) claims of Google's
\topics{} proposal in the context of realistic user behavior. Through new
statistical models of the distribution of user behaviors and resulting targeting
topics, we analyze the capabilities of malicious advertisers observing users
over time and colluding with other third parties. Our analysis shows that even
in the best case, individual users' identification across sites is possible, as
0.4\% of the 250k users we simulate are re-identified. These guarantees weaken
further over time and when advertisers collude: 57\% of users with stable
interests are uniquely re-identified when their browsing activity has been
observed for 15 epochs, increasing to 75\% after 30 epochs. While measuring that
the \topics{} API provides moderate utility, we also find that advertisers and
publishers can abuse the \topics{} API to potentially assign unique identifiers
to users, defeating the desired privacy guarantees. As a result, the inherent
diversity of users' interests on the web is directly at odds with the privacy
objectives of interest-disclosing mechanisms; we discuss how any replacement of
third-party cookies may have to seek other avenues to achieve privacy for the
web.
\end{abstract}

%% Keywords. The author(s) should pick words that accurately describe % the work
%being presented. Separate the keywords with commas.
\keywords{Targeted advertising, interest-disclosing mechanisms, Privacy Sandbox,
Topics API, cross-site tracking, third-party cookies}

\maketitle{}

\section{Introduction}\label{introduction}

Third-party cookies (TPCs), the historical \textit{interest-disclosing
mechanism} for online advertising, have been repeatedly shown to come at the
expense of user privacy with invasive tracking across websites. Deprecated or
soon-to-be by various web
actors~\cite{perry_design_2018,wilander_full_2020,brave_ok_2020,mozilla_firefox_2022,schuh_building_2019,schuh_buildingbis_2020},
different organizations have been proposing privacy-preserving alternatives that
could ultimately contribute to building a more private web for all. We refer to
the majority of these proposals as interest-disclosing mechanisms that assign
categories to users and disclose them to advertisers (\eg{} Google with
\floc{}~\cite{google_wicgfloc_2021,dutton_what_2021, google_federated_2021} and
\topics{}~\cite{dutton_topics_2022, google_github_2022}) or to other third
parties considered trusted (\eg{} \texttt{SPARROW} from
Criteo~\cite{criteo_sparrow_2020} or \texttt{PARAKEET} from
Microsoft~\cite{microsoft_github_2021}).

However, proposals that rely on interest-disclosure may be privacy-incompatible
with the natural diversity of user web behaviors that they relay. As a result,
user privacy provided by this type of proposals in the context of realistic user
behavior distributions remains unclear. We seek to evaluate the privacy and
utility guarantees of these proposals using as exemplar the \topics{} API from
Google--currently one of the most mature alternative interest-disclosing
mechanism to TPCs that Google plans to gradually deploy to users starting July
2023 with Chrome 115~\cite{merewood_preparing_2023}. In \topics{}, the web
browser collects and classifies the websites visited by users into topics of
interest. The top visited topics are updated regularly and observed by
advertisers to select which ad to display. A central privacy claim of \topics{}
is: \textit{``the specific sites you’ve visited are no longer shared across the
web, like they might have been with third-party
cookies''}~\cite{google_topics_2022}.

In this paper, we show that the privacy and utility claims of an
interest-disclosing mechanism (e.g., \topics{}) are directly at odds with the
same properties of users' browsing interests that make them unique. We
demonstrate through the \topics{} API how the disclosure of user interests can
be leveraged to re-identify users across websites, effectively violating one of
\topics{}'s guarantees. On the other hand, for any proposal to see market
adoption, user information returned to advertisers must be sufficiently accurate
to yield profitable ad targeting. Often called \textit{utility} within the
privacy community, we measure how accurately \topics{} maps user interests to
their visited websites and show how the \topics{} API can be abused to alter
this mapping. As the \topics{} API is still in a development phase, our
evaluation is based on the latest version (at time of submission) from May 30,
2023\footnote{Commit hash:
\href{https://github.com/patcg-individual-drafts/topics/tree/24c87897e32974c1328b74438feb97bf2ec43375}{24c8789}}
of the proposal~\cite{google_topics_2022}.

Through an analytical and empirical evaluation of the \topics{} proposal, we
develop statistical models based on realistic user web behaviors and
corresponding topics of interest. We show that advertisers and publishers can
observe users who have stable interests and leverage the results returned by the
API for re-identification. Indeed, we observe a highly skewed distribution of
topics among the top 1M most visited websites from CrUX top
list~\cite{durumeric_cached_2023}: 1 topic appears on more than 18\% of the
websites, only 196 topics out of the 349 from the taxonomy appear on more than
100 websites, and 42 topics are never observed at all. Using this prior, we
propose a way to identify noisy topics (\ie{} those returned randomly by the API
to provide plausible deniability and k-anonymity), remove them, and use the
genuine ones to track users across websites. We evaluate this phenomenon from
the points of view of a single and different websites that observe API results
for users across a single and several epochs\footnote{The Topics API currently
processes browsing histories once per epoch of size 1 week. See \autoref{topics}
for more details.} (see \autoref{tab:information_disclosure} for an overview of
our results).

In our evaluation, we show that an adversary can identify (1) about 25\% of the
noisy topics on single websites in \textit{one-shot} scenarios (wherein only one
result from the API is observed). As user behaviors are stable across epochs and
advertisers record the results returned by the API, we find that (2) the noise
removal increases to 49\% for 15 epochs and to 94\% when 30 epochs are observed
in \textit{multi-shot} scenarios. The identification of the genuine topics of
each user lays the foundations for re-identification of cross-site visits. We
find that, contrary to the goal of preventing re-identification, (3) 0.4\% of
the 250k users we simulate are re-identified by 2 advertisers colluding across
websites in one-shot scenarios, and that 17\% of them can be re-identified with
higher likelihood than just randomly. In multi-shot scenarios, (4) 57\% of the
users are uniquely re-identified and an additional 38\% are matched better than
just randomly in 15 epochs, while for 30 epochs 75\% are uniquely re-identified
and the rest (25\%) with a higher likelihood. This appears to directly violate
the privacy goals proposed by Google with \topics{} over TPCs. On the utility
perspective, we see that (5) \topics{} is quite useful to advertisers. On
average, the \topics{} API returns at least 1 true topic aligned with user
interests in about 60\% of cases--assuming the API is used faithfully. We
further demonstrate (6) how carefully crafted subdomains can alter this accuracy
and be abused to potentially assign unique identifiers to users. This paper
shows that the privacy-preserving claims of \topics{} are directly at odds with
user behaviors on the Internet. Other approaches may need to be explored to
develop a truly privacy-preserving alternative to TPCs.

We make the following key contributions:
\begin{itemize}
    \item We show how natural properties about user interests can break
    \topics{}'s privacy claims of non re-identification. Specifically, users
    with stable interests are as uniquely cross-site trackable with \topics{} as
    with TPCs.
    \item We find that \topics{} does not meaningfully lower the utility
    provided to advertisers from TPCs. We also identify ways to impact
    \topics{}'s privacy and utility if not used faithfully.
    \item We discuss how some mitigations to the \topics{} API can only be
    partial, and we point as well to other approaches than interest-disclosing
    mechanisms that may have to be sought for privacy-preserving online
    advertising.
\end{itemize}

\begin{table}
  \centering
  \caption{Overview of the risks of \topics{}'s information disclosure for
  different scenarios and cases of collusion (\autoref{privacy_eval}).}
  \label{tab:information_disclosure}
  \begin{tabular}{|c||c|c|}
    \hhline{-||--}
    \diagbox[width=\dimexpr\eqboxwidth{wd} + 5\tabcolsep\relax,
    height=0.8cm]{Collusion}{\raisebox{0.2ex}{Scenario}} & One-shot &
    \begin{tabular}[c]{@{}c@{}}Multi-shot \\(15-30 epochs)\end{tabular} \\
    \hhline{=::==}
    \begin{tabular}[c]{@{}c@{}}None \\\textbf{Noise removal}\end{tabular} &
    \begin{tabular}[c]{@{}c@{}}25\% of noisy \\topics removed \end{tabular} &
    \begin{tabular}[c]{@{}c@{}}49-94\% of noisy \\topics removed\end{tabular}\\
    \hhline{-||--}
    \eqmakebox[wd]{
      \begin{tabular}[c]{@{}c@{}}Across 2
      websites\\\textbf{Cross-site}\\\textbf{tracking}\end{tabular}}&\begin{tabular}[c]{@{}c@{}}0.4\%
      of users \\re-identified \\ 17\% better than\\
      just randomly \end{tabular} &
      \begin{tabular}[c]{@{}c@{}}57-75\% of users\\
      re-identified \\ 38-25\% better than\\
      just randomly\end{tabular} \\
    \hhline{-||--}
\end{tabular}
\end{table}
\section{Background \& Related Work}\label{background}
\label{relatedwork}

\shortsection{Third-Party Cookies \& Cross-site Tracking}
Web cookies, which offer websites the ability to record site-specific data in a
user's browser, are routinely abused to track users online. With TPCs,
advertisers can assign unique identifiers to web users, track them across
different websites, and obtain users' browsing histories. This is used to infer
user interests for targeted advertising~\cite{olejnik_why_2012,
karaj_whotracks_2019, bird_replication_2020, cook_inferring_2020}. As a result,
TPCs have been deprecated by different web actors (the Tor
Browser~\cite{perry_design_2018}, Safari WebKit~\cite{wilander_full_2020}, Brave
Browser~\cite{brave_ok_2020}, or Mozilla Firefox~\cite{mozilla_firefox_2022})
while others such as Google Chrome have announced their intention to do so in
the near future~\cite{schuh_building_2019,schuh_buildingbis_2020}.

\shortsection{Alternatives for Privacy-Preserving Advertising}
Deprecating TPCs altogether without offering any replacement would disrupt how
the ad-funded web presently operates. As a result, different organizations are
developing privacy-preserving alternatives for personalized advertising. The
focus of this paper as well as the majority of these proposals are based on
\textit{interest-disclosing mechanisms}. Generally, these solutions compute user
categories or assign each user to their interests. When advertisers and
publishers want to display an ad to users, that information is used to determine
which ad to show. The \floc{}
proposal~\cite{google_wicgfloc_2021,dutton_what_2021, google_federated_2021} and
later the \topics{} API~\cite{dutton_topics_2022, google_github_2022}, made by
Google as part of
\sandbox{}~\cite{google_privacy_2019,chavez_introducing_2022,google_privacy_2022},
assign users to a group of interests or classify user web histories into topics
categories, and then release those categories to advertisers through a web API
call. Two other proposals, \texttt{SPARROW} from
Criteo~\cite{criteo_sparrow_2020} and \texttt{PARAKEET} from
Microsoft~\cite{microsoft_github_2021}, introduce a trusted third
party--respectively, a gatekeeper and an anonymization service--to which user
data is disclosed to perform the ad selection process. On the other hand, a
different type of proposals for online advertising, like the \texttt{FLEDGE}
API~\cite{dutton_fledge_2022} from Google, assumes that user data should not
leave the browser and so executes the ad auction directly on users' devices.

\shortsection{Federated Learning of Cohorts (\floc{})} With \floc{}, an
alternative to TPCs developed by Google, participating web browsers weekly
compute the interest group (or cohort) their users belong to, based on their
browsing histories. Through a reporting mechanism to a central server, Google
ensures that the computed cohorts are either composed of enough users or merged
with other cohorts in order to provide some $k$-anonymity. Advertisers embedded
on visited webpages can observe user cohort IDs~\cite{dutton_what_2021,
google_federated_2021, google_wicgfloc_2021}. Analysis of \floc{} revealed a
variety of privacy concerns: (1) requirement in trusting a single actor to
maintain adequate $k$-anonymity, (2) concern that cohort IDs could create or be
linked to fingerprinting techniques, and (3) risk of re-identifying users by
tracking their cohort IDs over time and by isolating them into specific cohorts
through Sybil attacks
\cite{rescorla_technical_2021,berke_privacy_2022,turati_analysing_2022}. While
some parameters and details of \floc{} were still unclear, advertisers also had
concerns about how to interpret the cohort ID for utility. Google eventually
dropped \floc{} for the \topics{} API.

\shortsection{\topics{} API} \topics{} aims to replace TPCs for personalized
advertising. With this API, the web browser classifies the websites visited by
users into topics of interest. The top visited topics are updated once per epoch
and are observed by advertisers embedded on websites to select which ad to
display~\cite{google_topics_2022,dutton_topics_2022,google_github_2022}. See
\autoref{topics} for more details.

\shortsection{\topics{}
Analyses~\cite{epasto_measures_2022,thomson_privacy_2023,jha2023robustness}}
Along with its proposal, Google released a white paper analyzing the risk of
third parties re-identifying users across websites~\cite{epasto_measures_2022}.
First, an analytical evaluation is carried out to compute the aggregate
information leakage of \topics{} for two scenarios (per single and longitudinal
leakage) followed by an empirical experiment on a private dataset of synced
Chrome users browsing histories. The reported results show that the information
learned by a third party is somewhat limited compared to the worst case scenario
identified. This analysis is important for the discussion around \sandbox{}
proposals, but it also has limitations (some explicitly mentioned by the
authors): for instance, (1) it assumes that no actor is colluding with each
other when in practice advertisers could easily have such incentive, (2) some
uniform assumptions about the distribution and observations of topics are made,
(3) results are reported in aggregate potentially hiding risks for specific
users, (4) the noise in the mechanism is very briefly discussed, and (5) only 2
epochs were considered in the empirical evaluation. Our analysis of \topics{}
explicitly addresses these limitations through a realistic threat model, a more
thorough analysis over time (30 epochs), and a focus on the privacy consequences
of the diverse nature of user interests.

Following an inquiry from Google on their position about the adoption of the
\topics{} API~\cite{noauthor_request_2022}, Mozilla has released a privacy
analysis~\cite{thomson_privacy_2023} that points at shortcomings of \topics{}
and of the re-identification evaluation of Google's white paper. Thomson, the
author of this analysis, crafts a specific example of one user exhibiting a
unique interest among a population, to show the risk of being re-identified
through the \topics{} API. As proposed, a population as small as 70 users would
readily leak more information than the upper bound computed by Google. Thomson
additionally critiques the use of aggregate statistics, highlighting that
privacy guarantees must not only be assumed on average across web users but also
for individuals. In this paper, we analytically and empirically demonstrate the
actual consequences on the \topics{} API of the diverse nature of user
interests. We find that the distribution of topics among the top 1M most visited
websites is highly skewed, and use this information to identify some of the
noisy topics returned by \topics{}. By simulating 250k users across 30 epochs,
we demonstrate and quantify the risks identified in our analysis. Finally, we
measure the utility of the proposal for advertisers, a missing aspect from all
previous analyses on \topics{}.

In concurrent work, Jha et al., studied the privacy risk of re-identifying users
across websites through the Topics API for a substantially smaller simulated
population in a limited analysis~\cite{jha2023robustness}; while we perform a
broader, complete, and systematic analysis of both the privacy and utility goals
stated by Google on the proposal. Jha et al., collect data on a few real users
(\SI{268}{} in total) to simulate for the majority of their analysis a
population of \SI{1000}{users}; we propose a new methodology that directly uses
results from measurement studies on \textit{``several hundred million users''}
and representative of \textit{``over 95\% of page loads on the
Internet''}~\cite{ruth_world_2022,ruth_toppling_2022,durumeric_cached_2023,bird_replication_2020,olejnik_why_2012}
to simulate a population of \SI{250}{k} users. As a result, Jha et al., only
classify about 51k unique websites and observe a total of 250 topics; we
classify 1M websites for each CrUX and Tranco top-list observing 307 and 311
topics, respectively. Additionally, we systematically craft 3.5M adversarial
subdomains to study the potential for API abuse. Finally, if Jha et al.,
conclude that the attack seems impractical because it would require several
weeks; we argue that a possible re-identification attack appears at direct odds
with Google's stated goals. We show that some users are re-identified without
having to wait several epochs, not to mention that the size of an epoch could
change and be shortened in the future\footnote{As discussed here:
\url{https://github.com/patcg-individual-drafts/topics/issues/119}}.

\shortsection{A World Wide View of Browsing the World Wide
Web~\cite{ruth_world_2022}} Ruth et al., collaborated with Google to access a
private dataset about real browsing histories of Chrome users
worldwide~\cite{ruth_world_2022}. They were able to extract interesting
statistics and details about user browsing behaviors (some previously
conjectured by the community) that we directly use in our empirical analysis.
Their results show that web users always visit the same small number of websites
(25\% of page loads in their dataset come from only six websites with 17\% from
one website only) and spend most of their time on very few of them (10 websites
capture half of users' time spent online). They find that the top 10k and top 1M
most visited websites capture respectively from 70\% to 95\% of user traffic,
which justify using these rankings as a proxy to study users web browsing, even
though a lot of websites are visited relatively little which skews the analysis
towards the tail. Their results show that browsing behaviors tend to be similar
across regions for top use cases: users visit websites of similar categories
(search engine, video platforms, social networks, pornography, \etc{}), they
also explain that smaller populations and individuals exhibit different and
sometimes unique behaviors. Indeed, geographic, cultural, and linguistic
differences are observed, and so not every unique user web behavior may be
represented through global ranking lists.

\shortsection{Users Have Stable (and Unique) Web
Behaviors~\cite{tauscher_how_1997,montgomery_identifying_2001,kumar_characterization_2010,goel_who_2012,tyler_large_2015,muller_understanding_2015,hutchison_analysis_2005,mcdaniel_system_2006,olejnik_why_2012,bird_replication_2020}}
Multiple studies have been carried out since the early Internet area to measure
and evaluate user web behaviors. In the early 2000s, analyses identify how users
revisit websites~\cite{tauscher_how_1997} and what browsing trends and patterns
they exhibit~\cite{montgomery_identifying_2001}. These early studies already
find that user browsing behaviors and interests are stable over time, subsequent
studies come to the same observation. For instance, Yahoo! shows that webpages
of certain types and categories are revisited by the same user over
time~\cite{kumar_characterization_2010,goel_who_2012}. Studying search logs from
Bing, Microsoft finds that users exhibit consistent and stable domain
preferences over time, even during periods that would disrupt users' daily life
(like vacations), and that third parties have the ability to observe these
preferences~\cite{tyler_large_2015}. Diary studies such as Google's on the use
of tablet and smartphone devices also highlight that users have a diverse and
yet fixed set of activities they tend to perform
repeatedly~\cite{muller_understanding_2015}. On top of being stable, user
browsing behaviors have also been demonstrated to be unique: communities of
interests are used to re-identify
users~\cite{hutchison_analysis_2005,mcdaniel_system_2006}, and browsing
histories are shown to be unique by Olejnik et al.~\cite{olejnik_why_2012} and
in the replication study performed by Mozilla a few years
after~\cite{bird_replication_2020}.
\section{Exploring \topics{}}\label{topics}

In this paper, we use \topics{} as a canonical example of an interest-disclosing
mechanism, as it is currently the most mature proposal to replace TPCs. Our goal
is to analyze its privacy protections for users and its utility for ad-funded
websites. See \autoref{notations} for notations.

\subsection{\topics{} in Detail}
\label{topics_in_detail}

\begin{figure}
  \centering{}
  \includegraphics[width=.9\linewidth]{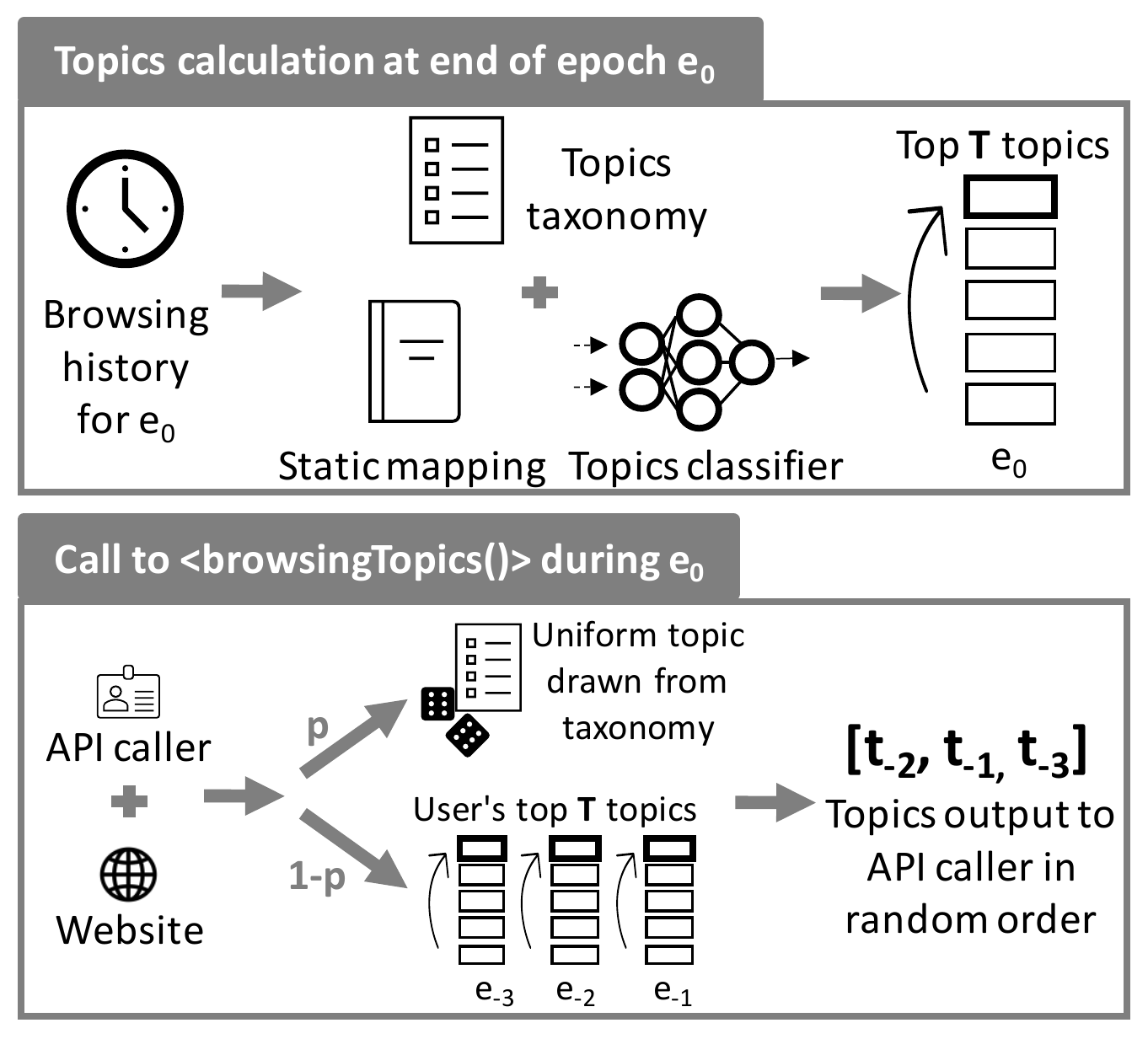}
  \caption{Overview of \topics{}.} \Description{The figure summarizes how the
  Topics API works. It depicts how at the end of an epoch, the calculation of
  the top topics of a user is done by considering the browsing history in that
  epoch and classify each visited hostnames into a topic from the taxonomy by
  using the static mapping and machine learning model released by Google. The
  figure then illustrates how a call to the Topics API by an advertiser embedded
  on a website determines the array of 3 topics that is returned; with some
  probability p, a random topic is drawn uniformly from the taxonomy, and with
  the opposite probability 1-p, it is picked from the top topics of the user for
  the corresponding epoch. It also shows that the returned array is randomly
  shuffled.}
  \label{fig:topics}
\end{figure}

With \topics{}, at the end of an \textit{epoch} $e_0$ (of size 1 week in the
current proposal) the browser--which globally tracks user's history--classifies
visited hostnames in order to compute the \textit{top} $T= 5$ \textit{most
visited topics}, which represents user interests. The initial \textit{taxonomy}
is composed of $\Omega=349$ different topics. To compute topics, the browser
first checks if the hostname is present in a \textit{static mapping} of
${\sim}10k$ most visited websites manually assigned to topics (if any). If not,
a \textit{machine learning classifier} is used (see \autoref{fig:topics}).
Hostnames are assigned from zero\footnote{For some hostnames, no topic is
returned because the classification is unknown or the website corresponds to
sensitive categories (ethnicity, sexual orientation, \etc{}).} to several topics
(one most of the time). Additionally, not all visited websites are taken into
account when computing a user's topics of interests in a given epoch $e_0$: only
the hostnames of the web pages that opted in to \topics{} and made a call to the
API are.

\textbf{API Call.} During epoch $e_0$, when publishers or advertisers embedded
on a web page call the \topics{} API, the browser will return them an
\textit{array of maximum $\tau = 3$ topics}: one per epoch before the current
epoch. For each epoch, the topic that is returned is either, with probability
$p=0.05$, a \textit{noisy topic} picked uniformly randomly from the taxonomy or,
with probability $1-p$, a \textit{genuine topic} picked randomly from the user's
top $T$ most visited topics for that epoch. These noisy topics are intended to
provide plausible deniability to users and ensure that a minimum number of users
is assigned to each topic (k-anonymity)~\cite{google_github_2022}. We study in
\autoref{privacy_eval} if these noisy topics can be identified by advertisers.
\topics{} also has a \textit{witness requirement} that ensures according to
Google that the \topics{} API does not disclose more information than
advertisers are already able to obtain with TPCs. With this witness requirement,
for advertisers to observe a genuine topic, advertisers must have already seen
that same topic on another website visited by the user in the previous $\tau$
epochs. If not, advertisers may be able to receive the parent topic of the
genuine one in the taxonomy, but only if they witnessed that parent topic in the
past as well. Additionally, if a topic is returned for a given epoch on a
website, any other subsequent call to the \topics{} API on that same website by
any caller will return the same topic for that epoch. Finally, advertisers may
not receive any topic; a user could have opted out of \topics{}, their web
browser does not support the API, they are in incognito mode, \etc{}

\textbf{Initial Taxonomy.} Google has released \topics{} with an initial
taxonomy of $\Omega=349$ topics~\cite{karlin_topicstaxonomy_v1md_2022},
seemingly curated from the taxonomy of \textit{Content Categories} of the Google
Natural Language Processing API~\cite{google_content_2022}. These topics are
alphabetically ordered and divided under 24 parent categories, \eg{} the
\textit{/Business \& Industrial} topic is a parent of \textit{/Advertising \&
Marketing}, that is itself a parent of \textit{/Sales} (see
\autoref{tab:taxonomy}). Additionally, Google removed topics that could be
deemed sensitive (ethnicity, sexual orientation, \etc{}).

\textbf{Static Mapping.} Google has released a list of manually annotated topics
for ${\sim}10k$ domains~\cite{dutton_topics_2022} that we refer to as the
\textit{static mapping}. Consisting of exactly 9254 domains,
\autoref{tab:static_countplot} shows the distribution of topics per individual
domain on this static mapping. The majority of these domains are assigned very
few topics (the median is 1 topic) and 1344 of them do not get assigned any
topic from the taxonomy at all, but instead the \unknownTopic{} topic (likely of
sensitive content).

\begin{table}
  \caption{Number of topics per individual domain in the static mapping of
  ${\sim}10k$ domain names annotated by Google.}
  \label{tab:static_countplot}
  \begin{tabular}{cc}
  \begin{tabular}{cc}
    \toprule
    \begin{tabular}[c]{@{}c@{}}Topic(s)\\ per domain\end{tabular} &
    \begin{tabular}[c]{@{}c@{}}Domains \\count\end{tabular}\\
    \midrule
    0 & \SI{1344}{} \\
    1 & \SI{4135}{} \\
    2 & \SI{2350}{} \\
    3 & \SI{1073}{} \\
  \bottomrule
\end{tabular}
&
\begin{tabular}{cc}
  \toprule
  \begin{tabular}[c]{@{}c@{}}Topic(s)\\ per domain\end{tabular} &
  \begin{tabular}[c]{@{}c@{}}Domains \\count\end{tabular}\\
  \midrule
  4 & 270 \\
  5 & 59 \\
  6 & 20 \\
  7 & 3 \\
\bottomrule
\end{tabular}
\end{tabular}
\end{table}

\textbf{Model Classifier.} Hostnames that do not appear in the static mapping
are classified through the use of a model that has been trained by Google. The
machine learning model of this classifier (weights, architecture, metadata,
\etc{}) is released publicly in the beta version of Google Chrome: Google uses a
Bert classifier~\cite{devlin_bert_2019} that accepts as input a string of
maximum 128 characters that has been tokenized and padded with spaces if
necessary. The output of the classifier is a vector of 350 confidence scores:
one for each of the 349 topics of Google's taxonomy to which an additional
\unknownTopic{} topic has been added. Although the model classifier used by
Google is public, its performance metrics such as accuracy and recall are not,
we fill that gap in \autoref{utility_eval} by evaluating the model classifier
performance. In order to exactly replicate the \topics{} API implementation from
Google Chrome, we identify the filtering applied to the output of the model by
Google: we detail this algorithm in \autoref{chrome_filtering_algo}. As a
result, we can directly reproduce the \topics{} API classification performed by
Google and classify any hostname we want: whether it is a real and registered
one like in \autoref{privacy_eval} when we classify the top 1M websites from
different top-lists, or a hostname that does not exist such as when we evaluate
if operators can influence the classification of their websites in
\autoref{crafting-subdomains}.

\textbf{Proposal Versions.} In this paper, we use the latest version of the
individual draft proposal of \topics{} from May 30, 2023 available on GitHub of
short sha commit
\href{https://github.com/patcg-individual-drafts/topics/tree/24c87897e32974c1328b74438feb97bf2ec43375}{24c8789},
the taxonomy
\textit{\href{https://github.com/patcg-individual-drafts/topics/blob/24c87897e32974c1328b74438feb97bf2ec43375/taxonomy_v1.md}{v1}}
of 349 topics, and the model classifier of version \textit{1} (used to be
labeled \textit{2206021246}). This corresponds to \topics{} version
\textit{chrome.1:1:2} being tested in Google Chrome beta at the time of
submission of this paper (May 2023). Google plans to start deploying the
\topics{} API gradually with Chrome 115 expected for July
2023~\cite{merewood_preparing_2023}.

\subsection{\topics{}'s Threat Model}

\begin{figure}
  \centering{}
  \includegraphics[width=.9\linewidth]{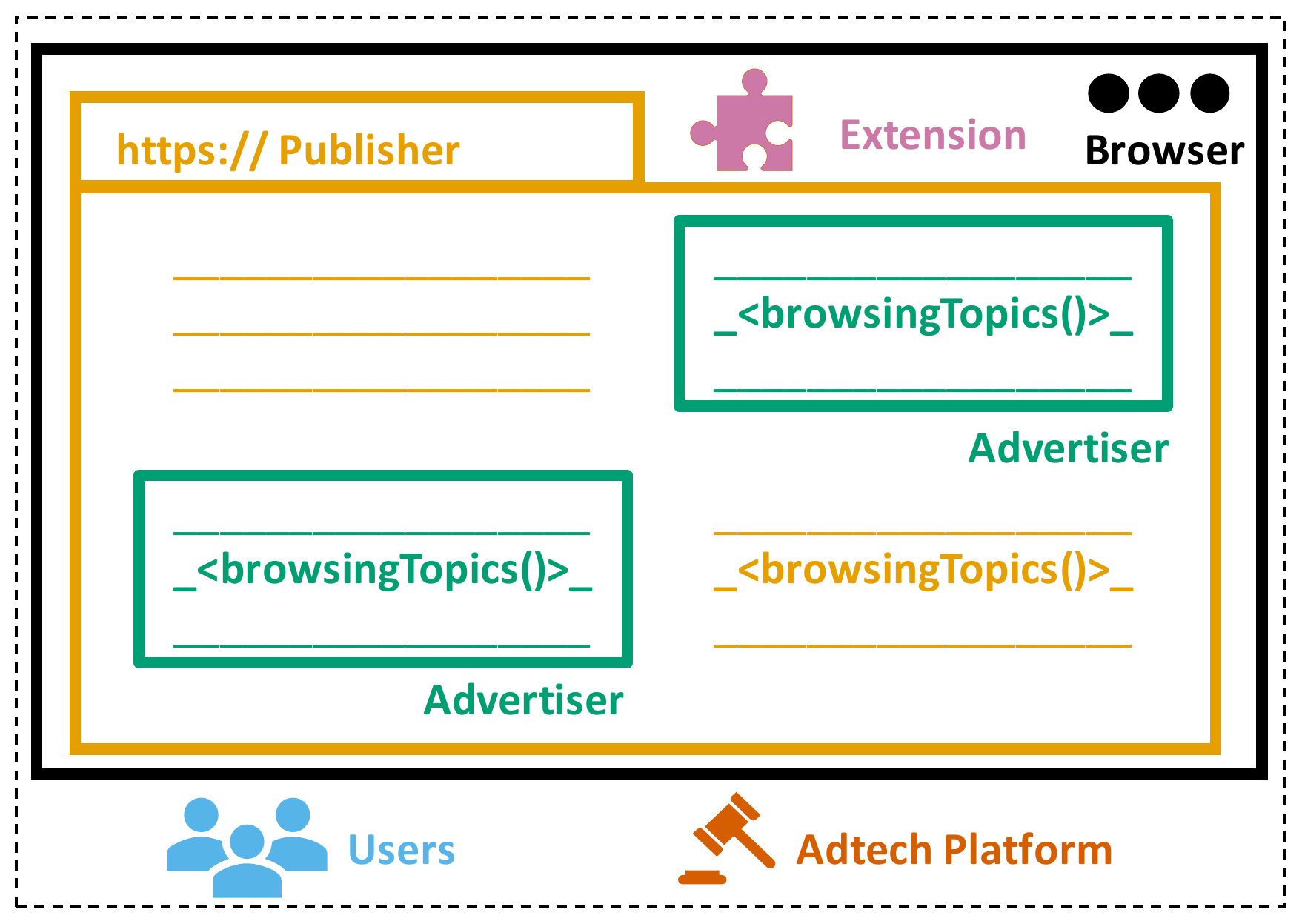}
  \caption{The different actors for \topics{} on a website's visit.}
   \Description{The figure illustrates the different actors present and involved
   when a user visits the website of a publisher; user(s), web browser, browser
   extension(s), publisher, advertiser(s), and adtech platform.}
  \label{fig:actors}
\end{figure}

Here, we present a realistic threat model for the \topics{} API (see
\autoref{fig:actors}). This model assumes \textit{users} accessing the content
of a \textit{publisher}'s website through their web \textit{browser}. On the
website along with the publisher's content, \textit{advertisers} embed scripts
to display ads to users after an ad auction was run on an \textit{adtech
platform}.

Under the \topics{} approach, advertisers can no longer use third-party cookies,
but they can call the \topics{} API to obtain user topics of interest (recall,
users are opted-in by default). While users and web browsers can be trusted in
that they faithfully follow the \topics{} protocol, a fundamental risk with
third parties is that they attempt to re-identify users across websites. The
threats are as follows: (1) advertisers will collude--there are strong
incentives for them to do so: better targeting users, improving their ad
selection, \etc{}--and (2) third parties (advertisers and publishers) will also
try to abuse the API--they can trick users into revealing certain topics by
clicking on specific URLs. Finally, even though we do not consider
\textit{extensions} in this paper (their current role and access to the
\topics{} API is unclear), we acknowledge that they may have to be part of the
threat model to assume by future work once \topics{} is deployed.

\subsection{\topics{}'s Privacy and Utility Goals}

The \topics{} proposal describes four goals across \textit{privacy},
\textit{utility}, and \textit{usability}. We next briefly discuss these goals
and our evaluation.

\begin{tcolorbox}
\goal{1} \textit{``It must be difficult to reidentify significant numbers of
users across sites using just the API.''}
\end{tcolorbox}
This is a \textit{privacy} goal; with \topics{} it should not be possible for
websites to identify that the same user visited them, as this would enable
cross-site tracking~\cite{olejnik_why_2012, karaj_whotracks_2019,
bird_replication_2020, cook_inferring_2020}. The phrasing used here is
ambiguous; it is not clear what \textit{``difficult''} and
\textit{``significant''} precisely mean in that context as they are not fully
defined. To perform our analysis, we define the \textit{difficulty} in breaking
user privacy to be the number of websites that the API caller needs to be
present on, or collude with, the number of topics that they need to observe, and
the needed number of observed epochs. For \textit{significance}, we measure the
proportion of $n$ users that can be re-identified, ideally and to be truly
private the \topics{} API should make this impossible for \textit{any} single
user, we quantify the re-identification risk of the \topics{} API in
\autoref{goal1-discussion}.

\begin{tcolorbox}
\goal{2} \textit{``The API should provide a subset of the capabilities of
third-party cookies.''}
\end{tcolorbox}

This is the \textit{utility} goal of \topics{}: the API should allow publishers
and advertisers to display targeted ads to the right users based on the returned
users' topic of interests. We evaluate how accurately browsing histories map
onto topics of interest in \autoref{utility_eval}.

\begin{tcolorbox}
\goal{3} \textit{``The topics revealed by the API should be less personally
sensitive about a user than what could be derived using today’s tracking
methods.''}
\end{tcolorbox}
This other \textit{privacy} goal mentions that \topics{}'s privacy disclosure
should leak less information about users than what could be inferred from TPCs
today. We analyze in \autoref{privacy_eval} if advertisers can denoise the
output of the API and re-identify users across websites.

\begin{tcolorbox}
\goal{4} \textit{``Users should be able to understand the API, recognize what is
being communicated about them, and have clear controls. This is largely a UX
responsibility but it does require that the API be designed in a way such that
the UX is feasible.''}
\end{tcolorbox}
The last goal mentioned by Google is about \textit{usability}; although it is
very important and should be taken into account when developing such an
API--especially if it were to be deployed to billions of internet users--we do
not consider this aspect in the rest of this paper. The reason is that a totally
different set of tools and expertise (\eg{} user studies, surveys, and
interviews) would be required than the ones we focus on to evaluate the privacy
and utility goals. We defer this usability evaluation to future work. The rest
of the paper evaluates \topics{} according to its privacy and utility goals.

\subsection{Information Disclosure}
\label{information_disclosure} \label{information_disclosure_noise}
\label{information_disclosure_tracking}

By returning user top interests, \topics{} discloses user information to
advertisers and alike. We now analyze the risks associated with this information
disclosure (see \autoref{tab:information_disclosure}) for different cases of
collusion between third parties (none and between advertisers across websites)
and scenarios within which the API was called: one-shot (wherein only one epoch
is observed per user) and multi-shot (several epochs observed per user). Recall
(\autoref{topics_in_detail}) that the disclosure of user interests by the
\topics{} API is \textit{limited}, \textit{noisy}, and its content
\textit{differs} across websites. However, users have stable web behaviors and
interests over time (see \autoref{relatedwork}), further amplified for their top
$T=5$ topics collected by their browser in the \topics{} API. As a result, we
must study the consequences of the stability of user interests on \topics{}'s
privacy claims.

\shortsection{No Collusion - Noise Removal} Consider the no collusion case: an
advertiser is embedded on a website and receives the topics of interest of the
users visiting it. A maximum of $\tau=3$ topics are observed per call. With a
probability $p=0.05$, each topic may be a noisy one picked from the taxonomy,
composed of $\Omega=349$ topics, instead of being one of the user's genuine
interests. This mechanism guarantees that for $n$ users visiting a website once,
an advertiser can expect to observe each topic in the taxonomy a minimum of
$\frac{n p \tau}{\Omega}$ times. Now, assume $N$ and $G$ the random variables
that count the number of noisy and genuine topics in an array of $\tau$ topics,
they have the following binomial distributions: $N \sim \mathcal{B}(\tau,p)$, $G
\sim \mathcal{B}(\tau,q=1-p)$. With the values from the current proposal,
advertisers can expect to get at least 2 genuine topics in 99.275\% of the
results that they observe in one-shot scenarios (where only one epoch is
observed per user). However, from just the outcome of this probabilistic
experiment, advertisers can not determine exactly which topics may be genuine or
noisy. Yet, they have a direct incentive to do so, for instance, to better
select which ad to display to user. This raises the question,
\textit{can third parties remove the noisy topics returned by the API?}

First, noisy topics are returned whether advertisers have observed them or not
for that user in the past epochs, \ie{} the witness requirement does not apply.
Advertisers who track the topics assigned to websites they are embedded on can
therefore easily flag noisy topics they do not have third-party coverage of.
Although, we can expect in practice that advertisers will be embedded on a large
set of websites as demonstrated by past measurement
studies~\cite{krishnamurthy_privacy_2009,mayer_third-party_2012,roesner_detecting_2012,acar_fpdetective_2013,acar_web_2014,libert_exposing_2015,englehardt_online_2016},
the distribution of topics on the most visited websites could inform advertisers
about which topics will appear more because they are noisy than genuine. Indeed,
we show in \autoref{distribution_mapping} that not all the topics from the
initial taxonomy are observed on the most visited websites, and build a
classifier to identify the noisy or genuine nature of topics.

Second, if a topic is repeated in the array of $\tau =3$ topics, advertisers can
distinguish between noisy and genuine topics. A topic that repeats $x$ times,
with $2\leq x \leq \tau$, would be noisy all these times with a probability of
$(\frac{p}{\Omega})^{x}$. By the opposite event rule, we have: a topic that
appears $x$  times is genuine at least once with probability
$1-(\frac{p}{\Omega})^{x}$, \ie{} more than 99.99\% for $x\geq2$. Users who have
stable interests across epochs have a higher chance of returning repeated topics
during a \textit{one-shot} scenario (\ie{} a single call to the API). Similarly,
advertisers in a \textit{multi-shot} scenario (\ie{} several calls to the API
are observed), have an incentive to collect user's interests across time. Doing
so, advertisers amass more information than through a single API call, and can
identify for instance a user's genuine topics when these repeat over
non-contiguous epochs or epochs separated by at least $\tau$ other epochs. As
users have stable topics (see \autoref{relatedwork}), their \topics{} API's
results across epochs on a website can be seen as a variant of the Coupon
Collector's Problem~\cite{motwani_randomized_1995}, and 11 epochs would be
necessary in expectation to see each one of the user's genuine topics once (see
\autoref{appendix:ccp} for proof and
\autoref{fig:multi_shots_min_median_max_top5} for empirical results from our
simulation). So, the more an advertiser observes a user across epochs, the more
confident it becomes in which topics are truly genuine or noisy; we further
study and quantify these risks in \autoref{noise_removal}.

\shortsection{Collusion - Cross-site Tracking} Advertisers may be able to remove
the noise added by the \topics{} API, especially for users with stable
interests. \textit{Can \topics{} be used to cross-site track users?}

During a given epoch, the \topics{} API returns a maximum of the same $\tau = 3$
topics to any caller embedded on a given website. For each consecutive epoch,
third parties that regularly call the \topics{} API are returned at most 1 new
topic per epoch. This effectively limits \topics{}'s privacy disclosure;
specifically, if user interests and their nature were uniform enough. However,
if we assume that the set of top $T =5$ topics for some user is stable, \ie{}
remains the same across epochs, a third party could potentially observe all top
$T=5$ topics of these users in as little as $T - \tau + 1 = 3$ epochs. Third
parties on other websites can do the same, and collude to re-identify users. The
initial taxonomy is composed of $\Omega=349$ topics, which leaves us with a
total of $\binom{\Omega}{\top} \approx 42~billion$ combinations of unique top
$T=5$ topics. Thus, if some users also exhibit unique interests, they risk being
re-identified across websites in both \textit{one-shot} and \textit{multi-shot}
scenarios. Also, even if users are sharing common interests with others, there
exists an arbitrary number of techniques out of scope of our threat model in
this paper that can be used on top of \topics{} to further discriminate users
into smaller and distinct
populations~\cite{hutchison_analysis_2005,mcdaniel_system_2006,thomson_privacy_2023},
making the risk of being re-identified real for everyone (see
\autoref{partial-mitigations}).

For a given epoch and user, calls to the \topics{} API that originate from
different websites do not return the same results every time. This is an attempt
at making it harder for advertisers that are colluding to re-identify users
across websites through one-shot scenarios. However, in multi-shot scenarios
where advertisers record topics returned for each user across epochs, more
information is accumulated. This directly paves the way for re-identification
attacks grouping users by their top $T$ topics as demonstrated in
\autoref{cross-site-tracking}. The natural diversity and stability of users
interests here again conflicts with the privacy guarantees that \topics{}
intends to provide.

\section{Privacy Evaluation}\label{privacy_eval}

In \autoref{information_disclosure}, we find that advertisers can remove the
noisy topics returned by \topics{} in one-shot and multi-shot scenarios, and
discuss that if third parties are colluding, users risk being tracked across
websites. We seek to empirically demonstrate and evaluate these risks:

\textbf{Q1: To what extent can third parties identify noisy topics?}

\textbf{Q2: To what degree can users be tracked across websites?}

\subsection{Challenge}

\begin{figure*}[!ht]
    \centering
    \begin{subfigure}[b]{0.33\linewidth}
        \centering
        \includegraphics[width=\textwidth]{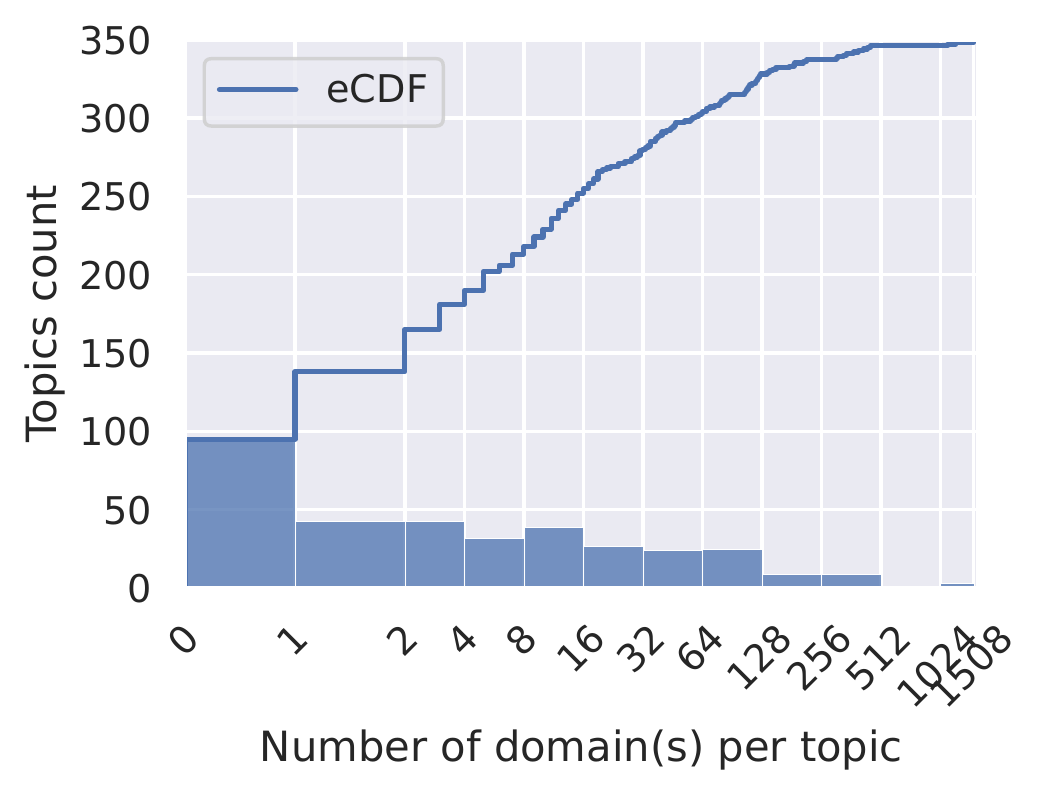}
        \caption{Static Mapping.}
         \Description{Distribution of number of individual topics per domain on
         the static mapping (histogram and cumulative distribution function).}
        \label{fig:static_distribution}
    \end{subfigure}
    \hfill
    \begin{subfigure}[b]{0.33\linewidth}
        \centering
        \includegraphics[width=\textwidth]{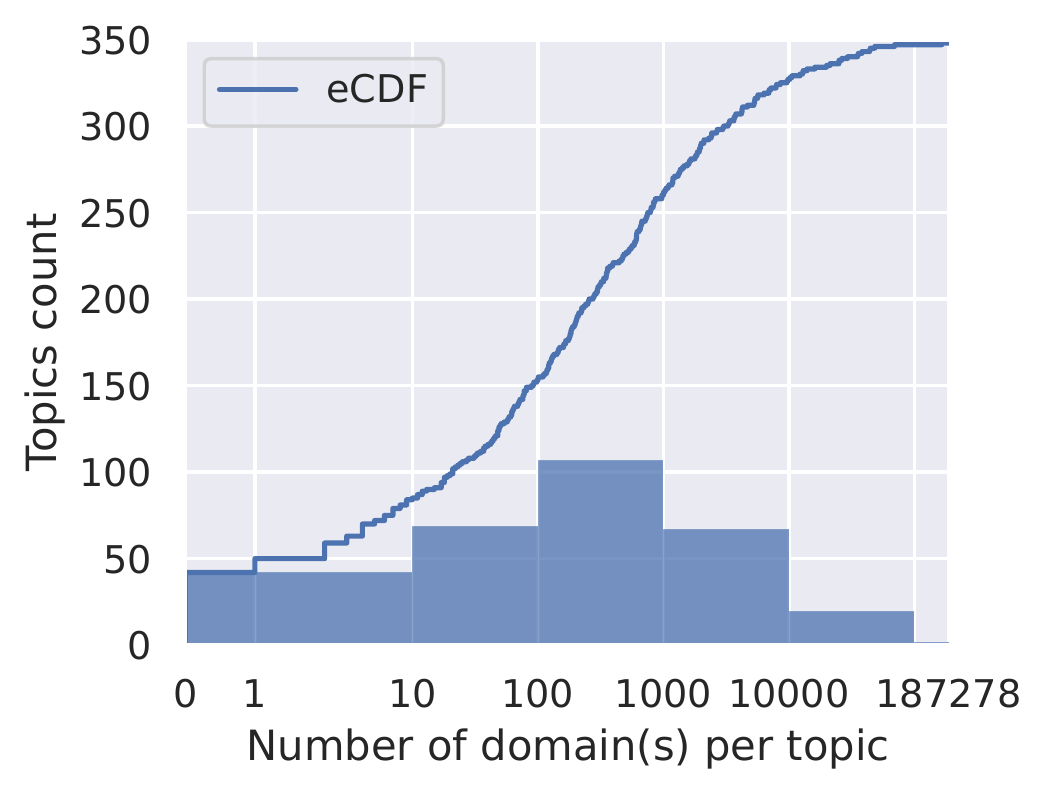}
        \caption{CrUX.}
        \Description{Distribution of number of individual topics per domain on
        CrUX (histogram and cumulative distribution function).}
        \label{fig:crux_distribution}
    \end{subfigure}
    \hfill
    \begin{subfigure}[b]{0.33\linewidth}
        \centering
        \includegraphics[width=\textwidth]{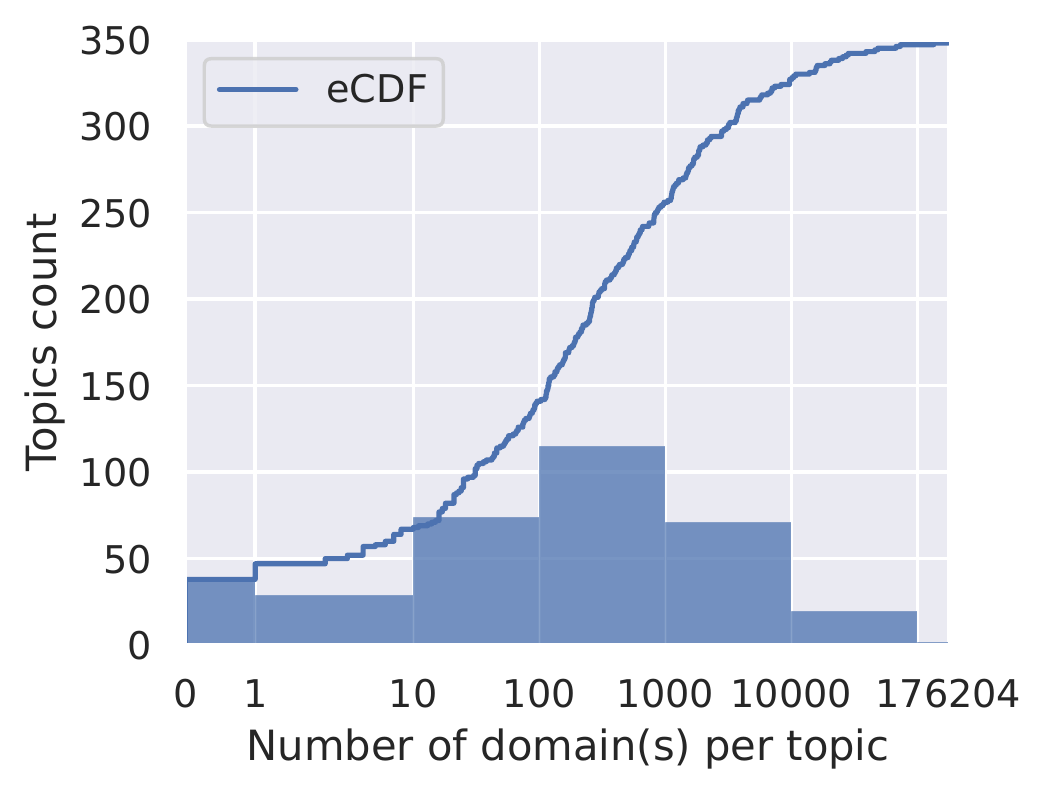}
        \caption{Tranco.}
         \Description{Distribution of number of individual topics per domain on
         Tranco (histogram and cumulative distribution function).}
        \label{fig:tranco_distribution}
    \end{subfigure}
    \caption{Distribution of the observations of each individual topics on the domains for each corresponding top list.}
     \Description{This figure is composed of 3 subfigures showing the
     distribution of number of individual topics per domain (histogram and
     cumulative distribution function) in: (a) the static mapping, (b) the CrUX
     top-list, and (c) the Tranco top-list. All figures show that some topics
     never appear at all on these list, some appear a few times only, and very
     few on a large set of domains from each list.}
    \label{fig:distributions}
\end{figure*}

To answer these questions through an empirical evaluation, it would be ideal to
have access to a recent and representative dataset of real browsing histories on
which the \topics{} API could be simulated. Unfortunately, no such dataset is
publicly available to researchers. Web actors, like Google, who collect this
browsing data at a large and systematic scale through opt-in telemetry and
reporting programs, keep it
private~\cite{google_chrome_nodate,bird_replication_2020}. Some online data
brokers do offer to sell some browsing histories datasets, but, for privacy,
ethical, and representativeness reasons about the unclear and vague collection
methodology of these datasets we immediately discard this possibility (see also
our ethics statement in \autoref{ethics}).

As a result, researchers have historically taken a survey approach to directly
ask users about their browsing habits and collect their
histories~\cite{tauscher_how_1997,montgomery_identifying_2001,kumar_characterization_2010,
goel_who_2012,muller_understanding_2015}. However, we observe that such
collection process is cumbersome and very often results in limited size of
collected samples for which ethical and representativeness questions still
arise. Not to mention that these researchers usually can not publicly release
their sensitive dataset of collected browsing histories, preventing others from
reproducing their results or methodologies without going through the same
collection process. Recognizing this challenge and aware of recent and
representative results published in the measurement community about online users
behaviors, we propose a new approach to solve this dataset problem for our use
case.

\subsection{Drawing from Representative Distributions}
\label{synthetic-dataset-construction}

First, we observe that to analyze the \topics{} API, we do not specifically need
detailed and timestamped browsing history traces but only the distribution of
the most visited domains for each user. Indeed, as explained in
\autoref{topics_in_detail}, the \topics{} API classifies the websites visited by
each user during an epoch into topics and keeps only the top 5 most visited
topics. Thus, we propose an alternative that lets us generate synthetic datasets
of any arbitrarily size. These are drawn directly from representative
distributions of online browsing behaviors aggregated on the large private
datasets of browsing histories that Cloudflare, Google, and Mozilla have
collected. Contrary to the datasets, these distributions are publicly available:
they have been published in measurement works performed in collaboration with
these organizations (see
\autoref{background})~\cite{bird_replication_2020,durumeric_cached_2023,ruth_toppling_2022,ruth_world_2022}.

Specifically, Mozilla reported the distribution of the number of unique domains
visited by 52k real users in a week in a replication study about the uniqueness
of browsing histories~\cite{bird_replication_2020}. Ruth et al., partnered with
Google and Cloudflare to perform a large-scale measurement of real users
browsing patterns of \textit{``several hundred million users
globally''}~\cite{ruth_world_2022,ruth_toppling_2022,durumeric_cached_2023}.
They disclosed the shape of the global distribution of web traffic that we use
on the top 1M most visited websites from the CrUX top-list, as it is
representative of \textit{``over 95\% of page loads on the
Internet''}~\cite{ruth_world_2022}. Note how the CrUX top-list is generated by
the same research group~\cite{durumeric_cached_2023}.

We will now walk through the generation of a synthetic dataset of a specified
size, before discussing different properties and advantages of our approach. We
sample a population with the same distribution of unique domains visited each
week by user as the one reported by Mozilla. Then, to determine which domains
are visited by each synthetic user, we use the global distribution of web
traffic reported by Ruth et al. This requires to first set a total order among
the top 1M websites of the CrUX top-list, which are binned by top rank (top 1k,
5k, 10k, 50k, 100k, 500k, and 1M). Fortunately, Ruth et al., report that they
see \textit{``Google, Youtube, Facebook, WhatsApp, Roblox, and Amazon within the
top six sites for at least ten countries''}~\cite{ruth_world_2022}, so, we use
the main FQDN of these organizations (\eg{} [\textit{www}
subdomain].[organization's name]. [\textit{com} global top level domain]) for
the top 6 websites. For the rest, we set a relative order within each bin by
using the Tranco rank~\cite{LePochat2019} of the \textit{eTLD+1} of each website
in CrUX. Similarly, we also use the ordered list of the top 100 \textit{eTLD+1}
globally returned by Cloudflare Radar's Domain Ranking
API~\cite{cloudflare_radar_2022}. This total ordering allows us to directly
sample browsing histories for each user according to their number of unique
visited domains.

Our approach has several advantages: (1) it does not require the collection or
use of any sensitive data by researchers, (2) the generated histories have the
same desired properties of representativeness as the global distributions see on
the web by Cloudflare, Google, and Mozilla because they are directly drawn from
their reported results, and (3) it allows the generation of any arbitrarily
large synthetic dataset than can (4) be publicly shared to (5) enable
reproducible methodologies. As such, we release as an \textit{open-source
artifact} \footnote{Available at
\url{https://github.com/yohhaan/topics_analysis}} the entirety of our generation
code and of our privacy and utility analyses of the \topics{} API performed for
this paper~\cite{beugin_topics_2023}.

\subsection{\topics{} Simulator}

For the purpose of our analysis, we implement a simulator to replicate what the
\topics{} API would output to different advertisers embedded on several websites
when the simulated population of users visit them. This simulator follows the
exact steps specified in the proposal of the \topics{}
API~\cite{google_topics_2022,dutton_topics_2022,google_github_2022}, with the
exception that we assume that advertisers have a large third-party coverage of
the web, thus, they can observe any topic from the taxonomy for every user. This
effectively removes the witness requirement of the \topics{} API and assumes a
worse case threat model that tends to be also more realistic: some advertisers
already have such important coverage as demonstrated by several past
studies.~\cite{krishnamurthy_privacy_2009,mayer_third-party_2012,roesner_detecting_2012,acar_fpdetective_2013,acar_web_2014,libert_exposing_2015,englehardt_online_2016}.

For the rest of this section, we generate two synthetic datasets following the
procedure  explained in \autoref{synthetic-dataset-construction}. These
histories are classified into the topics that would be returned by the \topics{}
classifier. As we are missing frequency information about the number of visits
to each unique website, we sample up to 10 sets of possible top $T=5$ topics of
interests among the topics observed, in case less than $T$ topics were observed
for a user, we draw the remaining ones uniformly from the taxonomy like the
\topics{} API does. We also assume that users have stable interests across time
(see related work cited in \autoref{background} and our discussion in
\autoref{discussion}). Finally, we present in \autoref{tab:datasets_stats} some
statistics about the two generated synthetic datasets used in the rest of this
section: (1) \textit{52k users} used to fine-tune our binary classifier in
\autoref{noise_removal}, and (2) \textit{250k users} on which we simulate the
\topics{} API for 30 epochs for 2 different advertisers to evaluate the
re-identification risk associated with \topics{}. Note that the generated
datasets have the same desired properties of representativeness 

\begin{table}
  \caption{Statistics of generated datasets for one epoch per user.}
  \label{tab:datasets_stats}
  \begin{tabular}{ccc}
    \toprule
    Metric& 52k users & 250k users\\
    \midrule
    Number of users & \SI{51977}{} & \SI{249997}{}\\
    Unique observed domains & \SI{91144}{} & \SI{248201}{} \\
    Unique observed topics & \SI{342}{} & \SI{349}{} \\
    Unique top $T=5$ profiles  & \SI{47928}{} & \SI{209745}{} \\
  \bottomrule
\end{tabular}
\end{table}

\subsection{No Collusion and Noise Removal}
\label{noise_removal}\label{distribution_mapping}

In this section, we analyze the possibility for advertisers to identify and
remove the noisy topics returned by the API. Recall that advertisers have an
incentive in doing so to improve, for instance, the selection of relevant ads to
display to users. As discussed in \autoref{information_disclosure}, if the
distribution of topics on the most visited websites is non-uniform, advertisers
can use it as prior information to discriminate topics on their genuine or noisy
nature. So, we ask here: \textit{what is the distribution of topics on the most
visited websites on the Internet?} and
\textit{how can we leverage that information in the context of the \topics{}
API?}

\shortsection{Topics Distribution as a Prior}
In \autoref{fig:distributions}, we plot the histogram and the empirical
cumulative distribution of how many individual topics are observed per number of
classified domains for (a) the static mapping, (b) the top 1M most visited
websites\footnote{CrUX provides a list of the top 1M most visited origins which
represents 991 656 unique domains in practice (some appear twice depending on
the \textit{http(s)} protocol used)} from
CrUX~\cite{durumeric_cached_2023,ruth_toppling_2022,ruth_world_2022}, and (c)
the top 1M most visited \textit{eTLD+1} from Tranco~\cite{LePochat2019}
classified with the \topics{} API.

The results show that the distributions of observations of each topic is very
non-uniform. First, the number of unique domains on which each topic is observed
tends to be rather small: the median is of only 3, 66, and 189 unique domains
per topic respectively for the classifications of the static mapping, CrUX, and
Tranco. Then, we observe that a moderate number of topics from the initial
taxonomy are never observed at all: 95, 42, and 38 topics respectively. Finally,
a few topics are seen on a significant number of domains: 3 topics are seen on
more than 10\% of the static mapping, when 1 topic in particular (\textit{\/Arts
\& Entertainment}) is seen in the classification of \SI{187 278}{domains} on
CrUX (18.8\%) and of \SI{176 204}{domains} on Tranco (17.6\%). Given that the
list of top 1M most visited websites represents ``95\% of all page loads on the
internet'' \cite{ruth_world_2022,ruth_toppling_2022}, the distribution of topics
among users is highly skewed as well.

Using this distribution as a prior, third parties can build a binary classifier
to distinguish between the noisy and genuine nature of topics. The first
approach is to set a global minimum number of websites among the top most
visited websites on which topics must appear to be considered genuine. Other
more advanced options that we do not explore here could see advertisers adapting
their strategies to the website they are embedded on, to the population
observed, or to other biases and signals, \etc{} To determine the minimal number
of websites a topic must appear on to be considered as genuine, we fine-tune our
simple binary classifier on the simulation of the \topics{} API for 1 epoch on
the small dataset of 52k users exclusively generated for that purpose. We
present in \autoref{tab:roc_52k} the raw results of the classifier for the
following thresholds: 0, 1, 2, 5, 10, 20, 50, 100, 500, and 1k websites, the
positive class of our classifier corresponds to the noisy nature of the topic
observed, while the negative one to genuine. For a threshold of 10 websites, we
observe that our classifier has still a very high true positive rate (TPR) for a
better false positive rate (FPR) than smaller thresholds. As a result, for the
rest  of our analysis, we set the global minimum threshold for the classifier to
10 websites, we also now exclusively simulate the \topics{} API on the larger
dataset of 250k users for all the others denoising and re-identification
experiments.

\begin{table}
    \caption{Results of our classifier for different thresholds. Highlighted row is the threshold used in the rest of the paper.}
    \label{tab:roc_52k}
    \begin{tabular}{ccccc}
      \toprule
      Threshold&Accuracy &Precision & TPR& FPR\\
      \midrule
      0 & 0.956& 0.130& 0.938& 0.044\\
      1 & 0.957& 0.144& 0.915& 0.043\\
      2 & 0.959& 0.173& 0.905& 0.041\\
      5 & 0.959& 0.207& 0.894& 0.040\\
      \rowcolor{Gray}
      10 & 0.961& 0.246& 0.900& 0.038\\
      20 & 0.956& 0.284& 0.681& 0.038\\
      50 & 0.958& 0.364& 0.658& 0.033\\
      100 & 0.959& 0.442& 0.619& 0.028\\
      500 & 0.912& 0.644& 0.315& 0.020\\
      1000 & 0.872 & 0.752& 0.246& 0.015\\
    \bottomrule
  \end{tabular}
  \end{table}

  \begin{figure*}
    \centering
    \begin{subfigure}[b]{0.33\linewidth}
        \centering
        \includegraphics[width=\linewidth]{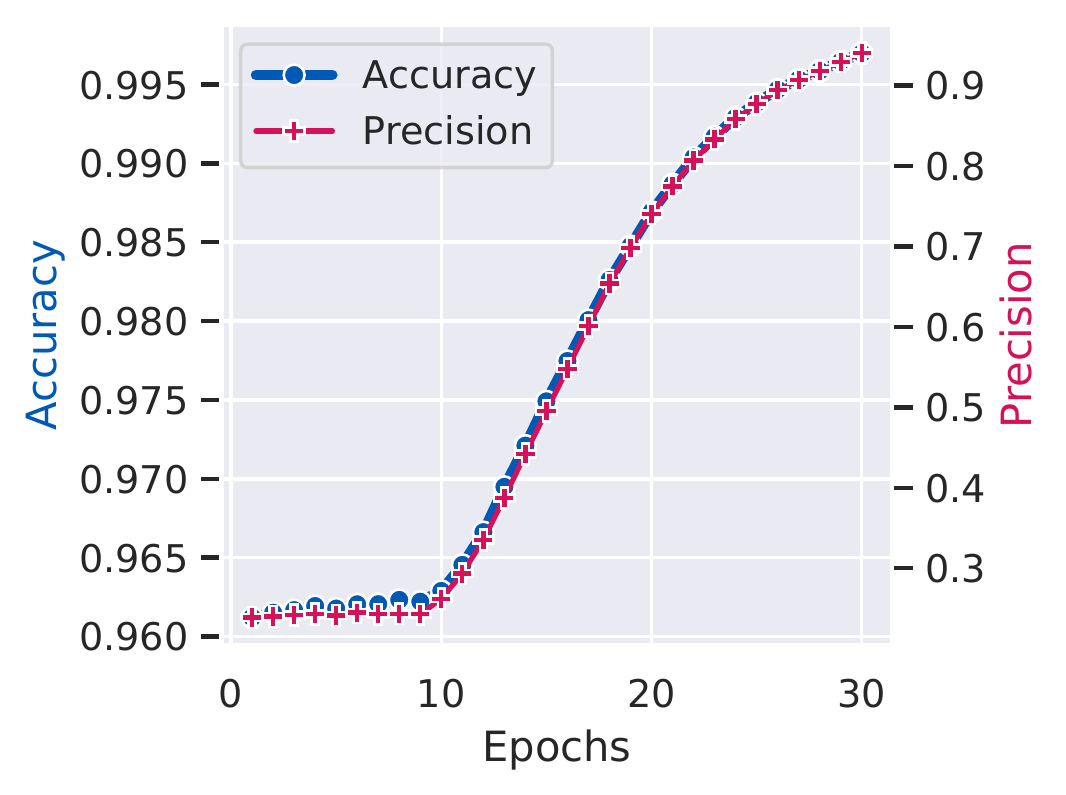}
        \caption{Classifier Accuracy and Precision.} \Description{Evolution
        across time of the accuracy and precision of our classifier that
        identify noisy topics. Accuracy goes from about 96\% to more than 99.5\%
        between epoch 1 and 30, while precision increases from less than 30\% to
        more than 90\%.}
        \label{fig:multi_shots_accuracy}
    \end{subfigure}
    \hfill
    \begin{subfigure}[b]{0.33\linewidth}
        \centering
        \includegraphics[width=\linewidth]{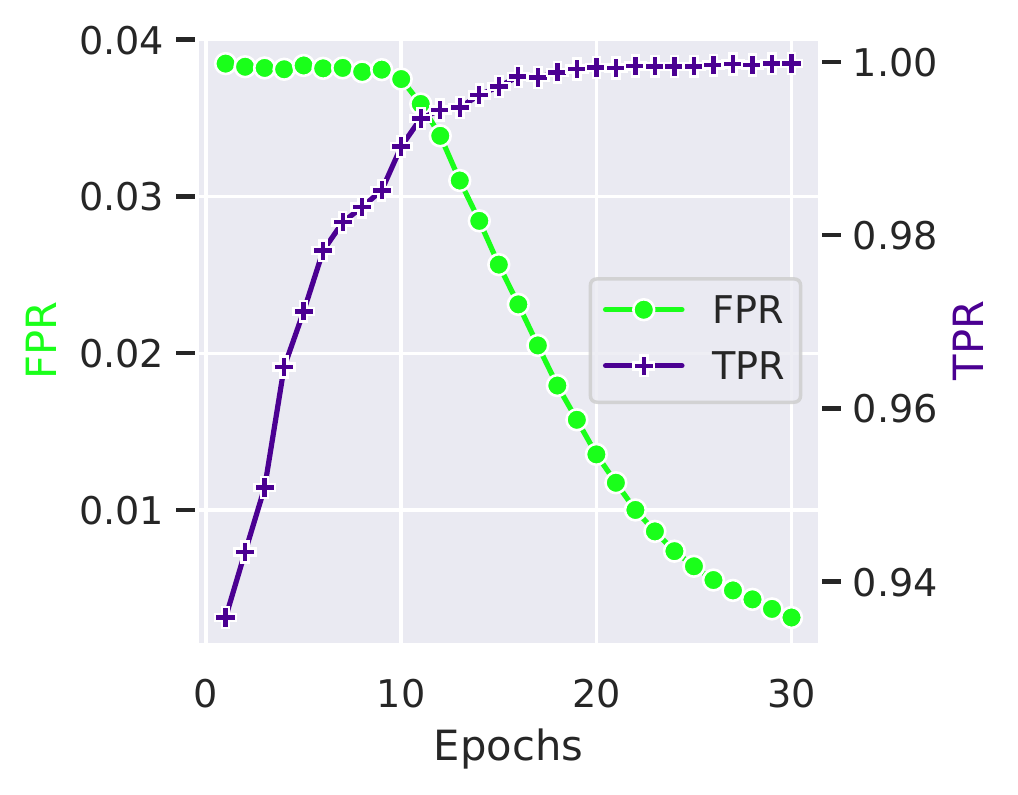}
        \caption{Classifier TPR and FPR.} \Description{Evolution across time of
        the False (FPR) and True (TPR) Positive Rates of our classifier that
        identify noisy topics. We observe that the FPR remains quite at the same
        level slightly below 4\% for 10 epochs before starting to drop to 0.3\%
        for 30 epochs, when the TPR goes from 94\% to 100\% between epoch 1 and
        30, at epoch 10 TPR is already above 99\%.}
        \label{fig:multi_shots_tpr}
    \end{subfigure}
    \hfill
    \begin{subfigure}[b]{0.33\linewidth}
        \centering
        \includegraphics[width=\linewidth]{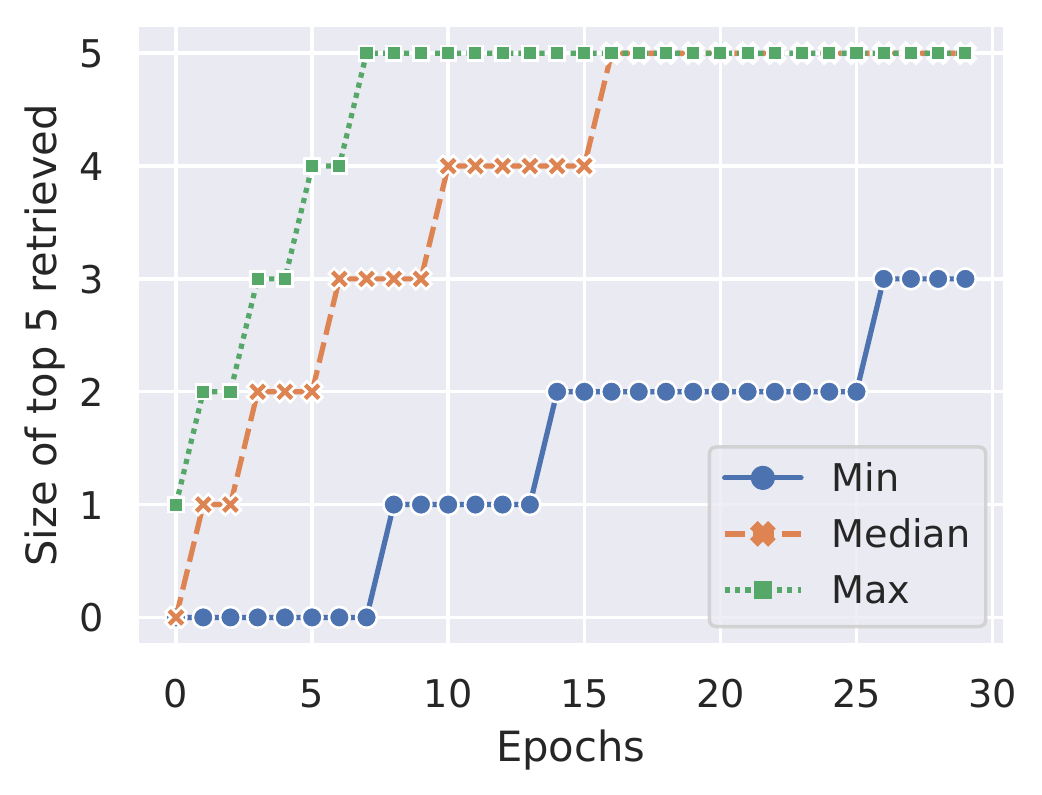}
        \caption{Min, Median, Max size of top 5 retrieved.}
        \Description{Evolution of the number of genuine topics retrieved from
        the top 5 of each user across epoch, we report the minimum, median, and
        maximum size retrieved for the 250k users over time. This is a step
        function, here are the values for the median: 0 topic retrieved after 1
        epoch, 1 topic retrieved for epochs 2 and 3, 2 topics retrieved for
        epochs 4 to 6, 3 topics retrieved for epochs 7 to 10, 4 topics retrieved
        for epochs 11 to 16, and the full top 5 retrieved in median starting
        from epoch 17.}
        \label{fig:multi_shots_min_median_max_top5}
    \end{subfigure}
    \caption{Multi-shot noise removal results for 250k stable users simulated across 30 epochs.}
    \Description{Figure composed of 3 subfigures presenting our results for
    multi-shot noise removal on the 250k stable users across 30 epochs.}
    \label{fig:multi_shots}
  \end{figure*}

\shortsection{One-shot Scenario}
In the one-shot scenario, advertisers only observe 1 result returned by the API
in a given epoch, \ie{} $\tau =3$ topics per user. Even though this information
disclosure is very limited, a topic that repeats in the returned array of a user
is way more likely to be genuine than noisy as explained in
\autoref{information_disclosure_noise}. We use this fact and our binary
classifier to measure how many of the noisy topics observed through the API can
be flagged as such by advertisers. For that, we simulate an advertiser that
observes the results returned by only one call to the \topics{} API for our
population of 250k users, \eg{} a total of 750k topics observed, from which
37.5k are expected to be noisy. 

For each user, the simulation returns $\tau=3$ topics only. Our classifier
determines which ones from these topics are likely genuine or noisy. First, if a
topic repeats among the topics of a user, it is flagged as genuine (recall that
in \autoref{information_disclosure_noise} we showed that it is very rare for a
topic to repeat if it is noisy). In the case of no repetition, we check if each
topic from the $\tau=3$ that have been observed is present on more than 10
websites on the classification of the top 1M websites from the CrUX top-list,if
so, it is classified as genuine, if not noisy. This one-shot procedure is
individually performed on-the-fly for each one of the 250k users. The results of
our noise removal mechanism are as follows: an accuracy of 96.1\%, with a
precision of 24.7\%--higher than the 95\% and 0\% a naive classifier always
outputting genuine would have achieved--, a true positive rate (TPR) of 93.9\%,
and a false positive rate (FPR) of 3.8\%. Thus, we find that our classifier
successfully identifies about 25\% of the noisy topics in one shot-scenarios.

\shortsection{Multi-shot Scenario}
Here, we are interested in the multi-shot scenario wherein advertisers record
across epochs the topics they observe for every user. We simulate a call to the
\topics{} API at every epoch for 30 epochs for every one of the 250k users.
Similarly to before, advertisers wanting to distinguish between noisy and
genuine topics should save the results returned for each user across time in a
first-party context. Then, they will check for repetitions within epochs and
across non-consecutive epochs, as observed earlier in
\autoref{information_disclosure_noise}; if topics repeat over non-consecutive
epochs or epochs separated by at least $\tau$ other epochs, we consider them
genuine. So, in the multi-shot scenario, we first attempt to identify genuine
topics; if we are able to recover top $T=5$ stable topics for each user, we then
mark all other topics as noisy. In the case where not enough repetitions have
been observed, we use the same binary classifier as in the one-shot
scenario\footnote{Note that when only 1 epoch is observed this multi-shot noise
removal is equivalent to the one-shot noise removal presented before.}. This
procedure is performed individually for each user and on-the-fly, \ie{} only
with all the topics observed during the epochs up to the one being simulated.

In \autoref{fig:multi_shots}, we respectively plot the evolution across 30
epochs of the accuracy, precision, TPR, and FPR of our multi-shot noise
classifier as well as the minimum, median, and maximum size of top $T=5$ genuine
topics that are retrieved across the 250k users. As expected; the more epochs an
advertiser observes results from the \topics{} API, the more confident it
becomes in which topics correspond to genuine or noisy ones as shown by the
evolution of the different metrics of our classifier across time. For instance,
if 93.9\% of the noisy topics are correctly identified for the first epoch, this
bumps to more than 99\% after 10 epochs, and almost reaches 100\% for 30 epochs.
Notice the change of trends observed around 10 and 11 epochs for different
metrics as we reach in expectation the number of necessary epochs to recover the
top topics of users as demonstrated in \autoref{appendix:ccp}. Ultimately, we
are recovering almost all the top $T=5$ genuine topics of stable users. Note
that advertisers being able to do so, prevents users from claiming plausible
deniability of being interested in some topics, which was what Google expected
to provide by adding these noisy topics as per their privacy goal \goal{3}.

\subsection{Collusion and Cross-site Tracking}
\label{cross-site-tracking}\label{goal1-discussion}

One of \topics{}'s privacy goals is to prevent third parties from being able to
re-identify users across websites \goal{1}. Here, we assume that 2 advertisers
$A$ and $B$ are colluding and sharing the topics they observe on 2 websites
$w_A$ and $w_B$ they are respectively embedded on. $n$ users visit these 2
websites at every epoch and the advertisers are trying to re-identify a portion
of or all users across these two websites. We ask:
\textit{how many users are they able to re-identify?}

We empirically measure this cross-site tracking risk by simulating the views
across 30 epochs that these 2 advertisers calling the \topics{} API would
observe for our 250k users. Note that this is the same setting assumed by Google
in their white paper released along with \topics{}~\cite{epasto_measures_2022}:
advertiser $A$ gets access to all the topics observed for each user $\{u_{i,B}
\mid 1 \leq i \leq n \}$ by advertiser $B$ and advertiser $A$ attempts to match
the user $u_{j,A}$ for some given $j$, with $1 \leq j \leq n$, they have
observed on website $w_A$ with the correct $u_{j,B}$ seen on $w_B$. In practice,
if \topics{}'s goal \goal{1} is respected, advertiser $A$ should not be
successful with a probability higher than $\frac{1}{n}$ for each user, \ie{} no
better than a random guess. We assume that a given user observed on $w_A$ is
\textit{uniquely} re-identified when it is exactly matched to its correct
identity on the other website $w_B$. We say that a user has a \textit{higher
likelihood} of being re-identified than randomly if advertiser $A$ identifies a
group of users seen by $B$ of size $k$ that contains the target user and such
that $k < n$.

Applying the techniques presented in the previous section on noise removal, each
advertiser simulated in our evaluation filters on-the-fly the noise from the
observations of topics, keeps only the observed topics deemed not noisy, and
comes up with the topics that are the most likely to be in the top $T=5$ of each
user. For each epoch, users that visited website $w_A$ are then mapped to the
user(s) observed on website $w_B$ with whom they share the most genuine topics
in common. Note that in this cross-site re-identification experiment, the roles
of advertisers and websites $A$ and $B$ are interchangeable.

\begin{figure}
  \centering{}
  \includegraphics[width=.9\linewidth]{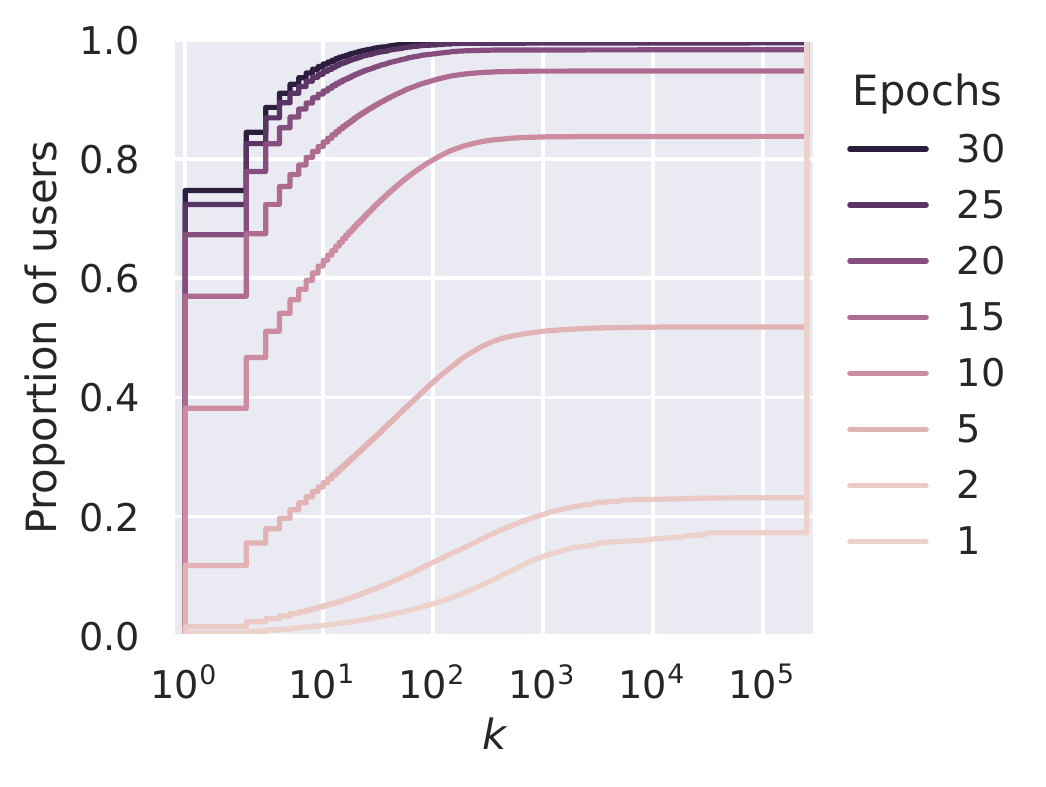}
  \caption{Distribution of the size $k$ of the re-identified groups.}
  \Description{On this figure, we plot different cumulative distribution
  functions of the proportion of the 250k users in our simulation that are
  re-identified to a group of size k users, i.e., the proportion of users that
  are provided with k-anonymity in our simulation. We plot the evolution of this
  distribution across 30 epochs, only epoch 1, 2, 5, 10, a5, 20, 25, and 30 are
  shown for readability. This figure shows that the more epochs are observed for
  each user, the more users are uniquely re-identified and the more users who
  are not uniquely re-identified are matched to groups of smaller sizes.}
  \label{fig:cdf_size}
\end{figure}

\shortsection{One-shot and Multi-shot Results}
In the one-shot scenario with collision, we find that 0.4\% of the 250k users we
simulate are uniquely re-identified and that 17\% of them can be with higher
likelihood than just randomly. Note that while the numbers obtained in one-shot
scenario are low, they are not null and some users (a total of 17.4\%) can be
re-identified across websites, which violates \topics's goal \goal{1} (modulo
the exact definitions of \textit{``difficult''} and \textit{``significant''}).

In multi-shot scenarios, the violation is even larger the more epochs are
observed: 57\% of the users are uniquely re-identified and an additional 38\%
with a higher likelihood than just randomly when 15 epochs are observed (for a
total of 95\% of the users), while 75\% and an additional 25\% of the users are
respectively re-identified uniquely and with a higher likelihood for 30 epochs
(total of 100\% of the users). These results across epochs are aligned with the
ones obtained when removing the noisy topics in the previous section; recall
that the more epochs an advertiser observes, the more genuine topics among
users' stable top $T=5$ they retrieved (
\autoref{fig:multi_shots_min_median_max_top5}), and as a result the more users
are re-identified, defying \topics{}'s goals \goal{1} and \goal{3}.

In \autoref{fig:cdf_size}, we now plot the cumulative distribution function for
different epochs of the proportion of users observed by advertiser $A$ across
each group size $k$ of re-identified users observed by advertiser $B$. As shown,
the proportion of uniquely re-identified users can be obtained for $k=1$, but
this graph also illustrates the evolution of the level of $k$-anonymity
(directly related to the size of the re-identified group) that a user in our
simulated population of size 250k users can expect across epochs. These results
directly inform us on how \textit{``difficult''} it is to re-identify
\textit{``significant numbers of users across sites''} \goal{1}, for instance:
for 10 epochs, over 60\% of the users can not be guaranteed strictly more than
10-anonymity in our evaluation.

\section{Utility Evaluation}\label{utility_eval}

We now evaluate the utility claim of \topics{}. Advertisers and publishers want
to serve ads that correspond to user interests to maximize the outcome (click,
visit, order, \etc{}). Thus, we ask:

\textbf{Q3: How accurate is the mapping of the \topics{} API between domains
visited by users onto topics of interest?}

We answer this by comparing the classifier accuracy from different approaches:
first, we measure the performance of the \topics{} model by comparing
classification results with the static mapping published by Google. While not
entirely confirmed, this static mapping likely constitutes a part of the
training or fine-tuning dataset for the \topics{} model. We then extend this
evaluation to the top 1M most visited websites using publicly available data on
site content. Finally, we look at the \topics{} model's resistance to
manipulation, evaluating the ability to craft subdomains that are misclassified
by the model.

\subsection{Static Mapping Reclassification}
\label{static-mapping-reclassification}

As explained in \autoref{distribution_mapping}, domains to be classified are
first checked against a static mapping of 9254 domains. If the domain is not
present in this static mapping, it is classified by the \topics{} model. We
first ask: \textit{does the model reflect human decisions?} We evaluate this
question by measuring the accuracy of the classifier on the static mapping that
was manually annotated by Google and so, can be considered as a form of ground
truth. After reclassifying these 9254 domains, we compute inferred topics for
each site using two different filtering techniques. First, we apply the same
filtering used by the Chrome browser (\textit{Chrome filtering}, see
\autoref{chrome_filtering_algo}). This filtering outputs a maximum of 3 topics
per domain, but the ground truth dataset has anywhere between 0 and 7 topics
associated with each domain. To allow for a best-case characterization of
\topics{}'s utility, we introduce a second filtering step that is more
conservative: \textit{top filtering} retains the same number of topics as seen
in the ground truth, giving \topics{} the best chance possible of matching
topics in the ground truth dataset.

\begin{table}
  \caption{Model performance on static mapping.}
  \label{tab:accuracies}
  \begin{tabular}{ccc}
    \toprule
    Metric&Top filtering & Chrome filtering\\
    \midrule
    Accuracy & 0.55 & 0.48 \\
    Balanced accuracy & 0.24& 0.23\\
    All topics correct (ratio) & 0.46 & 0.34\\
    Jaccard index (average) & 0.53 & 0.48\\
    Dice coeff. (average) & 0.56 & 0.52\\
    Overlap coeff. (average)& 0.56& 0.61\\
    At least one correct  (ratio)& 0.65 & 0.63\\
  \bottomrule
\end{tabular}
\end{table}

For each filtering strategy, we obtain two topic sets: a set of $\{predicted\}$
and $\{actual\}$ topics that we compare with different metrics reported in
\autoref{tab:accuracies} (see \autoref{notations} for formulas). Note here that
the difference we observe between accuracy and balanced accuracy can be
explained by the fact that most frequent classes contribute more to accuracy
than for balanced accuracy where each individual class's accuracy has an equal
weight computed by their recall~\cite{grandini_metrics_2020}.

Let's focus on the proportion of sets where all, some (Jaccard index, Dice
coefficient, and Overlap coefficient averages), or at least one predicted topic
are correct. These metrics show that at its best, the \topics{} model outputs at
least one topic in common with the ground truth on 65\% of the domains of the
static mapping. Note that Google did not disclose if this static mapping was
used to train or fine-tune \topics{}'s model classifier, though our results
would be broadly consistent with this. However, to understand how the \topics{}
model generalizes beyond potential training data, we next explore the
classifier's performance on a broader set of websites.

\begin{table*}
  \caption{Model performance when using Cloudflare Domain Intelligence API returned categories as ground truth.}
  \label{tab:cloudflare}
  \begin{tabular}{ccccccccc}
    \toprule
    Metric& Static Mapping & Top 1k & Top 5k & Top 10k & Top 50k & Top 100k &
    Top 500k & Top 1M\\
    \midrule
    Number of websites compared & \SI{2500}{} &\SI{423}{} & \SI{1977}{} &
    \SI{3880}{} & \SI{18160}{} & \SI{35466}{} & \SI{172529}{} & \SI{347686}{} \\
    Overlap coefficient (average)& 0.88 & 0.76 & 0.75 & 0.72 & 0.62 & 0.59 &
    0.55 & 0.53 \\
    At least one topic correct  (ratio) & 0.94 & 0.81 & 0.80 & 0.76 & 0.66 &
    0.63 & 0.59 & 0.57 \\
  \bottomrule
\end{tabular}
\end{table*}

\subsection{Top 1M Most Visited Websites}

To evaluate the model classifier of \topics{} more systematically, we ask
\textit{what would be the accuracy of the classifier on the most visited
websites?} Thus, we classify the top 1M most visited websites as reported by the
CrUX top list. We first manually verify a subsample of the classification and
then introduce a more systematic way to perform the comparison using one of
Cloudflare's APIs.

\shortsection{Manual Verification} For this manual verification, we are
interested in estimating the proportion of domains that get assigned to a valid
top topic, \ie{} \textit{is the topic assigned to the domain with the highest
confidence by the model related at all to the content of the corresponding
website?} For that, we perform a conservative sampling
approach~\cite{sampling_Thompson_3rdedition} and extract a uniform sample of 385
domains along their top topic from the classification of the CrUX top-list from
which we have excluded the domains from the static mapping (as already evaluated
in \autoref{static-mapping-reclassification}). The manual verification is done
by extracting the \textit{meta description} tag of each website and displaying
it along with the domain and classified topic. If this description can not be
obtained for various reasons (website unreachable, script blocking, \etc{}), we
manually get it from a Google search. Then, 3 persons independently evaluate if
the top topic for each domain is related to the meta description of the website
(translated to English with Google Translate if necessary). Finally, the results
are aggregated by keeping for each domain the most favorable rating, \ie{} a
domain is judged to have an invalid and unrelated top topic only and only if all
verifiers judged so. Through this population proportion
estimate~\cite{sampling_Thompson_3rdedition}, we find that the top label
returned by \topics{} is valid for 56\% of the domains. We also acknowledge that
this approach has its limitations: we only keep the top topic for each domain
when the other potential topics returned for that domain may be more accurate,
and the aggregation is biased, giving the benefit of the doubt to the \topics{}
API when one verifier judges its classification related to the meta description
of the website. To overcome these limitations, we next seek to automatically
compare the \topics{} API classification on the 1M top-list.

\shortsection{Cloudflare Categorization} To systematically evaluate the
\topics{} API on the top 1M domains from the CrUX top-list, we now propose to
compare the \topics{} classification to the content categories returned by the
Cloudflare Domain Intelligence API
\cite{cloudflare_cloudflare_2022,cloudflare_domain_2023}. These categories are
used in some of Cloudflare products to filter or block traffic based on certain
categories (security risks, adult themes, \etc{}).

We choose this service of Cloudflare Radar for different reasons: (1) it has
similar categories than Google's topics (facilitating the mapping between the
two), (2) it is a commercial system used in deployment by Cloudflare as part of
their Domain Intelligence and Threat APIs, (3) it aggregates different data
sources to perform the classification and only output categories on some domains
(more accurate than being just based on the hostname), (4) it is manually
curated (incorrect categories can be reported along with suggestions for domains
not classified yet), and finally (5) the API is accessible with a free account
(for reproducibility of some of our results)~\cite{cloudflare_glossary_2022}.

In order to compare Google's \topics{} classification to the categorization from
Cloudflare, we manually map \topics{}'s taxonomy to Cloudflare's 150 content
categories. First, we assign sensitive categories (Adult Themes, Drugs,
Religion, \etc{}) from the Cloudflare taxonomy to the \unknownTopic{} interest
from \topics{}, we then assign each of the remaining 349 topics to every content
category it could be mapped into. We further refine our assignment by looking at
the domains that do not correctly get their topics mapped to content categories
from the static mapping (which explains the high performance results on the
static mapping). Then, we categorize the top 1M most visited websites from the
CrUX top list with the Cloudflare API, and we keep only the domains for which
some content categories are returned. For each domain, we end up mapping these
categories to a set of topics that we can compare to the ones predicted by
\topics{}. We release our manual mapping as part of our
\href{https://github.com/yohhaan/topics_analysis}{open-source
artifact}~\cite{beugin_topics_2023}.

\autoref{tab:cloudflare} shows for different ranks of top most visited websites
the overlap coefficient average and the proportion of domains for which there is
at least one topic in common between both classifications, using the remapped
Cloudflare's output as ground truth. These results show that the \topics{}'s
classification is also quite accurate beyond the $\sim10k$ websites from the
static mapping; indeed on the domains from the top 1M that are categorized by
the Cloudflare API, at least 1 topic is output in common by \topics{} on 57\% of
the cases. We conclude that this matching implies moderate utility of the
\topics{} model for targeted advertising \goal{2}. In practice, user interests
are tagged heuristically based on visits to many sites, and the aggregation of
these visits would reduce classification mistakes and improve accuracy over
these per-site results.

\subsection{Crafting Subdomains}
\label{crafting-subdomains}

Motivated by a discussion on the proposal~\cite{karlin_should_2022}, we study
the privacy and utility trade-off of not allowing advertisers to set their own
topics. At present, the \topics{} model only uses the hostname of the websites
as input; this is limited information to work with compared to potentially
having access to some content of the website (such as the meta description for
instance) or having publishers providing their own topics or classification
hints. The purpose of this limitation is ostensibly to reduce the ability of
website operators to influence their site tagging (and thereby reducing API
abuse to assign unique identifiers to users). However, this also reduces the
utility of the system because many hostnames are incorrectly classified. In this
section, we evaluate if this utility trade-off actually provides defense against
manipulation: \textit{can publishers influence the classification through the
use of carefully crafted subdomains?}

To demonstrate the extent to which this is possible, we carry out the following
experiment where we craft subdomains for each of the top 10k most visited
websites from CrUX. As a preliminary step, we classify with \topics{} every
individual word from the English dictionary provided by
WordNet~\cite{miller-1994-wordnet}. Then, for each domain in the top 10k, we
craft 350 subdomains by preprending to it the English word output with the
highest confidence for each topic. To illustrate, we provide the following
example: \textit{``batman''} and \textit{``dance''} are the two words classified
with the highest confidence across WordNet to the \textit{``Comics''} and
\textit{``Dance''} topics, respectively. Thus, for the domain
\textit{``example.org''}, we would craft the following subdomains:
\textit{``batman.example.org''} and \textit{``dance.example.org''},
respectively. We classify with \topics{}, this total of 3.5M new subdomains and
interpret the results through the following two types of misclassifications:
\textit{untargeted elimination} and \textit{targeted addition}. An untargeted
elimination would be used to eliminate an undesirable topic association from a
site, whereas a targeted addition could introduce a new, desirable topic
association. This could be done to improve the mapping, help publishers or
advertisers observing topics from websites that they would not be embedded on
otherwise, and to some extent alter the topics of some users (with a potential
privacy risk of assigning unique interests).

\begin{figure}
  \centering{}
  \includegraphics[width=.9\linewidth]{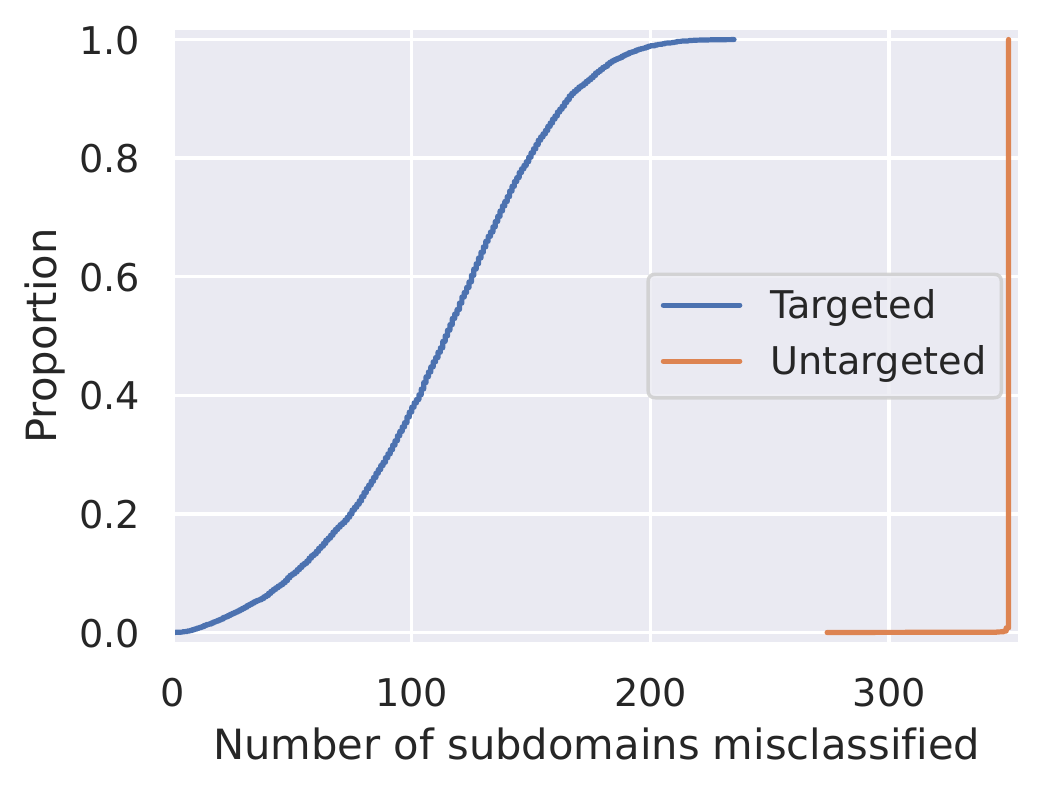}
  \caption{Cumulative distribution of the number of successful targeted
  additions and untargeted eliminations per domain.}\Description{Here we plot
  the cumulative distribution of the number of successful targeted additions and
  untargeted eliminations per domain. We see that almost for all the 10k
  original domains considered we can force a misclassification. In the targeted
  case, this plot shows that we are able to craft 114 subdomains that are
  misclassified as wanted for at least half of the 10k domains, with at most 235
  targeted subdomains crafted for some domain.}
  \label{fig:subdomains}
\end{figure}

\autoref{fig:subdomains} shows the cumulative distribution for the original
domains of the number of crafted subdomains that were respectively
misclassified. For untargeted elimination, we look at if the classification of
the newly crafted domain changed at all from the classification of the initial
domain; this happens for almost all the crafted subdomains for any domain in the
top 10k. We are also interested in a specific targeted addition, in this case,
we are successful on more than 114 crafted subdomains for half of the initial
10k domains, and at most we are able to successfully craft a total of 235
targeted subdomains. These results show that the model classifier is quite
susceptible to changes to the initial domain as untargeted elimination is
successful in almost all the cases. For targeted addition, while our approach is
quite simple, it is sufficient to target a fair number of topics. We are only
interested in showing that capability here, but one could improve these results
by applying for instances techniques from adversarial machine
learning~\cite{goodfellow_explaining_2015,papernot_limitations_2016,carlini_towards_2017,madry_towards_2022,sheatsley_space_2022}.

Our results demonstrate that publishers can craft specific subdomains to set
their own topic(s). They can implement this in practice by redirecting users to
them to influence the \topics{} computation. However, this is contradictory to
the security goal that domain-based classification gives up utility to achieve.
Using more site data to determine classification would further exacerbate this
issue. As is, \topics{} does not achieve an optimal trade-off between security
and utility, where restricting classification further (such as basing only on
\textit{eTLD+1}) would provide more security with minimal utility trade-off. As
the current system effectively allows sites to set their own topic, we argue
that such a feature should either be made openly available to publishers in an
accessible and easy-to-audit way to incentivize honest participation or further
restricted.
\section{Discussion}\label{discussion}

The privacy risks we observe with \topics{} arise from the API relaying the
underlying distribution of user interests. Users have diverse web behaviors that
are unlikely to change and so natural properties of their interests, such as
heterogeneity, stability, and uniqueness, are reflected through the API results.
These can be exploited as shown in \autoref{privacy_eval}, but parts of the
system can also be modified to try to make the \topics{} observations less
skewed and more uniform. We next present some partial mitigations and explain
why they do not fix the problem, we then discuss the different assumptions of
our analysis, and discuss future work and other directions such as other types
of approaches that may have to be considered to enable privacy-preserving online
advertising.

\subsection{Partial Mitigations}
\label{partial-mitigations}

Recall that in \autoref{privacy_eval}, we observe that the distribution of
topics is highly skewed on the top 1M most visited websites. A direct
consequence is that some topics are more likely to be noisy when observed by
advertisers than genuine. To attempt to fix this distribution, a new taxonomy
and classifier could be designed so that all the topics appear more uniformly
than they currently do on the top 1M websites. In this regard, several
modifications are possible: (1) a larger training dataset with observations of
all the classes, (2) extending the static mapping with observations for every
topic, (3) providing more information to the classifier than just the hostname
of the website (although this also introduces accuracy and privacy issues as we
saw in \autoref{crafting-subdomains}), (4) ensuring that every topic from the
taxonomy appears genuinely on the most visited websites, (5) removing altogether
topics that appear very little in practice (although this impacts accuracy for
users and specific websites from these categories), or (6) splitting topics that
appear a lot and merging the ones that appear less. However, for these
mitigations, a crucial assumption is made about which domain the observations
are made on. For instance, fixing the classifier and the taxonomy so that every
topic on the top 1M most visited websites appears a minimum of times, does not
imply that this would also be the case on the top 10k, top 100k, for the
visitors of a website about some given subject, or for the users of a smaller
group than the larger population created by advertisers based on additional
fingerprinting vectors (location, language, etc.). As a result, these would only
be partial mitigations, as they do not address the underlying diverse nature of
all user interests. Not to mention that they can directly impact the accuracy
and level of utility of the API.

\subsection{Assumptions of our Analysis \& Future Work}
\label{assumptions-future-work}

In order to perform our analysis, several assumptions were made. In this
section, we list them and discuss their foreseen implications.

\shortsection{Population Size}
The results presented in this paper are based on the simulation of a population
of 250k synthetic users, but our simulator can very well generate a population
of any size. Thus, we can evaluate the impact of the population size on the
re-identification results of our analysis. \autoref{tab:population_size_unique}
presents for different population sizes the proportion of users that are
uniquely re-identified across time. As expected, we can see that for a given
epoch the re-identification rate is higher the smaller the population is,
indeed, we simulate advertisers that are trying to recover the genuine top $T=5$
topics of interests for each user and the smaller the population is, the smaller
the probability is that other users have the same top interests. As a result,
this experiment also shows that the more unique the top interests of each user
are among a given population, the higher the risk that they are re-identified
across websites. Finally, if we assume that third parties can link the results
of the \topics{} API with other fingerprinting signals (outside the scope of our
analysis), such as location, language, etc., they can easily discriminate a
large population of users into smaller groups within which re-identification
becomes easier. Further work in that space is needed to evaluate the
consequences of such scenario.

\begin{table}
    \caption{Proportion of users uniquely re-identified for different population sizes.}
    \label{tab:population_size_unique}
    \begin{tabular}{cccccccc}
      \toprule
      Epoch & 1k & 5k & 10k& 50k& 100k& 250k& 500k\\
      \midrule
      1 & 0.115 & 0.045 & 0.032 & 0.012 & 0.008 & 0.004 & 0.004 \\
      2 & 0.192 & 0.101 & 0.076 & 0.037 & 0.026 & 0.015 & 0.011 \\
      5 & 0.509 & 0.362 & 0.312 & 0.199 & 0.160 & 0.116 & 0.095 \\
      10 & 0.792 & 0.709 & 0.674 & 0.530 & 0.467 & 0.382 & 0.337 \\
      15 & 0.874 & 0.847 & 0.831 & 0.726 & 0.676 & 0.567 & 0.536 \\
      20 & 0.900 & 0.899 & 0.896 & 0.822 & 0.784 & 0.670 & 0.654 \\
      25 & 0.912 & 0.919 & 0.917 & 0.861 & 0.831 & 0.722 & 0.711 \\
      30 & 0.916 & 0.927 & 0.924 & 0.878 & 0.851 & 0.745 & 0.737 \\
      \bottomrule
    \end{tabular}
\end{table}

\shortsection{Stability of Users Interests} In this paper, we assume users
interests to be stable across time, while related work have shown that users
have stable and unique web
behaviors~\cite{tauscher_how_1997,montgomery_identifying_2001,kumar_characterization_2010,goel_who_2012,tyler_large_2015,muller_understanding_2015,hutchison_analysis_2005,mcdaniel_system_2006,olejnik_why_2012,bird_replication_2020}
(see \autoref{relatedwork}), we also have to acknowledge that not every user
would exhibit such stability in practice. Our analysis can be seen as a
worst-case analysis regarding this point: users who do have stable interests
across time have thus a higher risk of being re-identified. Additionally, if
stability makes users more re-identifiable, another important factors, as
described in the previous paragraph, is the uniqueness of the top interests of
each user among the given population. We defer to future work the evaluation of
different stability scenarios of users interests, but our denoising and
re-identification results (\autoref{fig:multi_shots}, \autoref{fig:cdf_size},
and \autoref{tab:population_size_unique}) when only a few epochs have been
observed and just some of each user's interests retrieved can already inform us
that advertisers would still be able to re-identify a portion of the users.

\shortsection{Third-Party Coverage} For our analysis, we disregard the witness
requirement of the \topics{} API by assuming that advertisers have a large
enough third-party coverage of the top 1M websites. This can be considered as a
worst-case scenario, in practice, not all advertisers are embedded on all
websites. However, past measurement studies have shown that some advertisers
already have a very large third-party coverage of the
web~\cite{krishnamurthy_privacy_2009,mayer_third-party_2012,roesner_detecting_2012,acar_fpdetective_2013,acar_web_2014,libert_exposing_2015,englehardt_online_2016},
and as such other analyses of the \topics{} API take the same assumption as
well~\cite{epasto_measures_2022,thomson_privacy_2023,jha2023robustness}. We
leave to future work the measurement and evaluation of more realistic coverage
scenarios.

\subsection{Other Avenues}
\label{other-avenues}

By design, interest-disclosing mechanisms report user information to third
parties, other avenues or ideas to replace TPCs with a truly private solution
may be found in additional proposals of different nature. Other alternatives,
such as the \texttt{FLEDGE}~\cite{dutton_fledge_2022} and
\texttt{TURTLEDOVE}~\cite{google_wicgturtledove_2020} proposals, assume a
different setting wherein ad selection is done locally in web browsers without
user data ever leaving their devices. However, more work remains to be done to
evaluate these proposals. Additionally, building a more private web goes beyond
the replacement of just TPCs. Other open challenges include how to perform,
record, and communicate conversion and impression metrics in a private way that
still lets advertisers get the utility they would like. Google is for instance
leading the development of \sandbox{} initiative that, \textit{``aims to create
technologies that both protect people's privacy online and give companies and
developers tools to build thriving digital
businesses''}~\cite{google_privacy_2021}.

For the Web, Google's proposals also aim at preventing fraud and spam (Private
Stakes Tokens API~\cite{dutton_private_2021}), measuring ads conversion
(Attribution reporting API \cite{nalpas_attribution_2021}), and reducing
cross-site privacy exposure (First Party Sets~\cite{dutton_first-party_2021},
Shared Storage API~\cite{white_shared_2022},
CHIPS~\cite{mihajlija_cookies_2022}, etc.). For the Android mobile OS, the goals
are similar: reducing user tracking by deprecating access to cross-app
identifiers such as Advertising ID, and limiting the scope that third party
libraries in applications can
access~\cite{chavez_introducing_2022,google_sdk_2022,google_privacy_2022}. Other
organizations contribute to \sandbox{} or to their own projects, for instance:
Apple with Private Click Measurement~\cite{apple_inc_private_2021}, Brave with
Brave Private Search Ads~\cite{brave_brave_2022}, or Meta and Mozilla with
Interoperable Private
Attribution~\cite{taubeneck_interoperable_2022,mozilla_privacy_2022}. Work
remains to be done to seek and evaluate these other avenues.

\subsection{Ethics Statement}
\label{ethics}

In this paper, we chose for ethical reasons, privacy concerns, and
representativeness issues not to collect or acquire any dataset of real user
browsing histories. Instead, we use publicly available aggregated ranked lists
of top visited
websites~\cite{LePochat2019,google_chrome_nodate,durumeric_cached_2023,ruth_toppling_2022}
and rely on recently published results from measurements works about web user
browsing
behaviors~\cite{bird_replication_2020,ruth_world_2022,ruth_toppling_2022} to
generate synthetic browsing histories. We find it to be the only way to pursue
and evaluate these proposals' claims without having access to representative
browsing history datasets (\ie{} without being one of the large web actors) and
without sustaining the business model of data brokers (see
\autoref{privacy_eval}). We hope that by releasing our code as an open-source
artifact, we inspire others to adopt a similar methodology.

\section{Conclusion}\label{conclusion}

Several privacy-preserving alternatives like \sandbox{} are being developed at
the moment by web and online advertising actors. With the deployment of these
alternatives, the modifications being introduced could impact billions of users
and lead to a better web ecosystem--some proposals even aim at improving user
privacy beyond the web. However, this could also very well be for the worst, if
we replicate similar errors to the ones that were made in the past with the same
technologies that we are trying to replace today. As a result, it is of the
upmost importance to pay attention to the changes being proposed, analyze and
evaluate them, and attempt to foresee their potential consequences in the
context of realistic user behaviors. In this paper, we have taken on this
endeavor for interest-disclosing alternative mechanisms for online advertising,
such as Google's \topics{} API--currently aimed to be gradually deployed to
Chrome users starting July 2023 with Chrome 115--, and we have quantified how
their privacy objectives and design are directly at odds with the natural
diversity of user behaviors.

\nocite{Tange2011a}

\begin{acks}
We would like to thank Cloudflare for allowing us to classify the top 1M most
visited domains with their Domain Intelligence API by increasing the API rate
limits of our account. The authors also thank Eric Pauley, Ryan Sheatsley,
Kunyang Li, Quinn Burke, Rachel King, and Blaine Hoak for their feedback on
initial versions of this paper. We also sincerely thank the reviewers for their
insightful and constructive comments and suggestions on our paper.

\textbf{Funding acknowledgment:}
This material is based upon work supported by the National Science Foundation
under Grant No. CNS-1900873. Any opinions, findings, and conclusions or
recommendations expressed in this material are those of the author(s) and do not
necessarily reflect the views of the National Science Foundation.
\end{acks}

\bibliographystyle{ACM-Reference-Format}
\bibliography{refs}

\appendix{}
\section{Taxonomy}
\label{tab:taxonomy}

\begin{table}[!ht]
\caption{Distribution of topics in initial taxonomy; number of topics (including parent topic) per parent category.}
\begin{tabular}{lc}
\toprule
Parent category & Number of topics \\
\midrule
/Arts \& Entertainment & 56 \\
/Autos \& Vehicles & 29 \\
/Beauty \& Fitness & 14 \\
/Books \& Literature & 3 \\
/Business \& Industrial & 23 \\
/Computers \& Electronics & 23 \\
/Finance & 23 \\
/Food \& Drink & 8 \\
/Games & 16 \\
/Hobbies \& Leisure & 11 \\
/Home \& Garden & 8 \\
/Internet \& Telecom & 11 \\
/Jobs \& Education & 13 \\
/Law \& Government & 4 \\
/News & 7 \\
/Online Communities & 4 \\
/People \& Society & 9 \\
/Pets \& Animals & 9 \\
/Real Estate & 3 \\
/Reference & 4 \\
/Science & 10 \\
/Shopping & 10 \\
/Sports & 33 \\
/Travel \& Transportation & 18 \\
\bottomrule
\end{tabular}
\end{table}

\section{Comparison metrics used in model performance analysis}

When comparing the $\{actual\}$ and $\{predicted\}$ sets of topics, we report
the average across the domains considered for the following:

\[ \text{Jaccard\_index} = \frac{|\{actual\} \cap \{predicted\}|}{|\{actual\}
\cup \{predicted\}|} = \frac{\text{Dice\_coefficient}}{2 -
\text{Dice\_coefficient}}\]

\[ \text{Dice\_coefficient} = \frac{2|\{actual\} \cap
\{predicted\}|}{|\{actual\}| + |\{predicted\}| } =
\frac{2~\text{Jaccard\_index}}{1+\text{Jaccard\_index}}\]

\[ \text{Overlap\_coefficient} = \frac{|\{actual\} \cap \{predicted\}|}{
\min{(|\{actual\}|,|\{predicted\}|)} }\]

Note that always $\text{Jaccard\_index} \leq \text{Dice\_coefficient}$.

\section{Chrome filtering strategy}
\label{chrome_filtering_algo}

Google does not explain in the \topics{} proposal how the output of the model
classifier is filtered. We identify in the chromium source code the
component\footnote{\textit{src/components/optimization\_guide/core/page\_topics\_model\_executor.cc}}
that handles this filtering. As different set of parameters are used in the
corresponding chromium's unit tests, we verify that we have the correct
combination by validating that our classification of 1000 words chosen randomly
from the English dictionary matches (except for a negligible number of floating
point arithmetic issues) the one from Google --accessible from the
\textit{chrome://topics-internals} page in Google Chrome beta. Next, we explain
the algorithm that Google uses (also provided in natural language in
\autoref{alg:chrome_filtering_algo}); first the top 5 topics ordered by their
confidence score are kept, the sum of these 5 scores is computed, and any topic
whose score is lower than 0.01 is discarded. Then, if the \unknownTopic{} topic
is still present and if its score contributes to more than 80\% to the previous
sum, the domain is considered as sensitive; none of the 349 topics is returned,
and \unknownTopic{} is output. If the proportion is lower, the \unknownTopic{}
topic gets dropped from the top topics considered for the domain. In this case,
the scores of the remaining topics are normalized by the total sum computed
previously and if the result is bigger than 0.25, only then the topic is kept.
If after all of these steps, none of the 349 topics remain, then \unknownTopic{}
is returned.

\begin{algorithm}[!ht]
  \SetKwComment{Comment}{/* }{ */} \caption{Chrome filtering
  strategy}\label{alg:chrome_filtering_algo} \SetKwComment{Comment}{/* }{ */}
  \DontPrintSemicolon \KwIn{topics\_classified \ie{} the 350 topics and their
  score} \KwOut{predicted topics as \topics{} output in Chrome} \Comment{Chrome
  filtering parameters} $max\_topics \gets  5$\; $min\_unknown\_score \gets
  0.8$\; $min\_topic\_score \gets 0.01$\;
  $min\_normalized\_score\_within\_top\_n \gets 0.25$\; \Comment{Initialization
  of variables} $top\_sum \gets 0$\; $unknown\_score \gets 0$\; \Comment{Return
  highest $max\_topics$ top scores} $topics \gets
  sort\_return\_top(topics\_classified,max\_topics)$\; \For{$topic$ in
  $topics$}{ $top\_sum \gets top\_sum + topic.score$\; \If{$topic.id$ is
  \unknownTopic{}}{ $unknown\_score \gets topic.score$\; } }
  \Comment{\unknownTopic{}'s score is considered too strong} \If{$unknown\_score
  / top\_sum > min\_unknown\_score$}{ \Return{\unknownTopic{}}\; }
  $topics\_predicted \gets None$\; \For{$topic$ in $topics$}{ \Comment{Check
  minimal and normalized scores} \If{$topic.id$ is not \unknownTopic{}
  \textbf{and} $topic.score \geq min\_topic\_score$ \textbf{and} $topic.score /
  top\_sum \geq min\_normalized\_score\_within\_top\_n$}{
  $topics\_predicted.add(topic.id)$\; } } \If{$topics\_predicted$ is $None$}{
  \Return{\unknownTopic{}}\; } \Else{ \Return{topics\_predicted}\; }
\end{algorithm}

\section{\topics{} \& Coupon Collector's Problem}
\label{appendix:ccp}

In this section, we make the connection between the \topics{} API in the case of
users with stable interests across epochs and the Coupon Collector's Problem of
probability theory~\cite{motwani_randomized_1995}. We show how \topics{} is a
modified version of the classical Coupon Collector's Problem. In this problem, a
person collects coupons, cards, tokens, or other items obtained through a
probabilistic experiment such as a random drawing (buying gum packages, cereal
boxes, \etc{}). The collector seeks to get all the items at least once, or
several sets of each item, \etc{}

A user with stable interests across epochs has a top $T=5$ topics of interests
that is fixed. They are genuine and one of them is output with probability
$q=1-p=0.95$ for each individual epoch instead of a noisy random one.
Advertisers are interested in obtaining only the genuine topics to improve their
ad selection. As a result, in \topics{}, advertisers and alike would be the
collectors, and the topics observed the coupons (one new one per corresponding
epoch). As we assume users with stable interests here, this places us in the
Coupon Collector's Problem settings of a drawing with replacement.

Let's denote by $E$ the time (number of individual epochs) when all genuine $T$
topics have been collected (\ie{} observed at least once) by advertisers. If we
denote by $e_i$ the time at which the genuine topic $i$ was collected after
having collected $i-1$ other topics, we have; $ E = e_1 + e_2 + \ldots + e_{T}$

Collecting a new genuine topic $i$ after having collected $i-1$ other genuine
topics for a user has a probability of $q\frac{(T - i + 1)}{T}$. Thus, $e_i$
follows a geometric distribution of expectation $\frac{T}{(q)(T - i + 1)}$. And:

\begin{align*}
  \E(E) &= \E(e_1) + \E(e_2) + \ldots + \E(e_{T}) = \frac{T}{q} \left( 1 + \frac{1}{2} + \ldots + \frac{1}{T} \right)\\
  &= \frac{T}{q} \sum_{i=1}^{T} \frac{1}{i} \approx 13~\text{for $T =5$}
\end{align*}

On expectation, advertisers must observe the topics for a user that correspond
to 13 individual epochs to say that they collected all genuine topics of their
top $T=5$ topics. This means 11 consecutive API calls as a minimum to \topics{};
as the initial call discloses a maximum of 3 new topics and the consecutive ones
a maximum of 1 new topic at once.

\section{Notations}\label{notations}

\begin{table*}[!ht]
    \centering
    \caption{Notations and symbols used in this paper.}
    \label{tab:freq}
    \begin{tabular}{ccc}
      \toprule
      Notation&Definition&Value\\
      \midrule
      Commit on \topics{} proposal & Commit sha of the \topics{} proposal
      studied in this paper
      &\textit{\href{https://github.com/patcg-individual-drafts/topics/tree/24c87897e32974c1328b74438feb97bf2ec43375}{24c8789}}
      (May 30, 2023) \\
      Taxonomy version & Version of the taxonomy used in this paper & v1\\
      Model version & Version of the studied \topics{} classifier in this paper&
      1 (used to be labeled 2206021246)\\
      \topics{} version & Latest version of \topics{} studied in this paper
      &\textit{chrome.1:1:2} \\
      CrUX version & Version of the CrUX top-list used in this paper
      &\href{https://github.com/zakird/crux-top-lists/raw/main/data/global/202212.csv.gz}{202212}
      (December 2022) \\
      Tranco version & Version of the Tranco top-list used in this paper
      &\href{https://tranco-list.eu/download/6JZJX/1000000}{6JZJX} (February 6,
      2023) \\
      \browsingTopics{} & \topics{} API call &\textit{document.browsingTopics()}
      \\
      $A,B$ & Advertisers $A$ and $B$ & - \\
      $\mathcal{B}$ & Binomial distribution & - \\
      $e_i$ & Epoch $i$ & Size $e_{i+1}-e_i = 1~\text{week}$  \\
      $i,j$ & Generic math variables used for iterations & - \\
      $n$ & Number of users& -\\
      $p$ & Probability to pick a random topic from taxonomy & 0.05\\
      $q=1-p$ & Probability to pick a genuine topic from user's top $T$ topics &
      0.95\\
      $T$ & Number of top topics per epoch & 5 \\
      $\tau$ & Number of topics returned by \browsingTopics{} & 3 maximum \\
      $t_j$ & Topic $j$ & -\\
      TPCs & Third-party cookies & - \\
      $u_{i,A}$ & User of identity  $i$ observed by advertiser $A$& -\\
      $w_B$ & Website on which advertiser $B$ is embedded & -\\
      $\Omega$ & Number of topics in taxonomy & 349 topics (+ \unknownTopic{}
      topic)\\
    \bottomrule
  \end{tabular}
\end{table*}

\end{document}